%% file: PAT.tex
\def \VersionLong {}
\ifdefined\VersionLong
	\newcommand{\LongVersion}[1]{\ifdefined\VersionWithComments{\color{red!40!black}#1}\else#1\fi}
	\newcommand{\ShortVersion}[1]{\ifdefined\VersionWithComments{\color{black!40}#1}\fi}
\else
	\newcommand{\LongVersion}[1]{\ifdefined\VersionWithComments{\color{black!40}#1}\fi}
	\newcommand{\ShortVersion}[1]{\ifdefined\VersionWithComments{\color{red!40!black}#1}\else#1\fi}
\fi

\documentclass[conference]{IEEEtran}
\IEEEoverridecommandlockouts

\usepackage[utf8]{inputenc}
\usepackage[english]{babel}

\usepackage[ruled,vlined,linesnumbered]{algorithm2e}
	\SetKwInOut{Input}{input}
	\SetKwInOut{Output}{output}

\usepackage{subcaption}

\usepackage{paralist} %

\newenvironment{ienumeration}
	{\ifdefined\VersionLong\begin{enumerate}\else\begin{inparaenum}[\itshape i\upshape)]\fi}
	{\ifdefined\VersionLong\end{enumerate}\else\end{inparaenum}\fi}

\usepackage{amsmath} %
\usepackage{amssymb} %

\usepackage{eurosym}

\LongVersion{
	\usepackage[backend=biber,backref=true,style=alphabetic,url=false,doi=true,defernumbers=true,sorting=anyt,maxnames=99]{biblatex} %
	\addbibresource{pat.bib}

	\DeclareSourcemap{
		\maps[datatype=bibtex, overwrite]{
			\map{
			\perdatasource{Andre.bib}
			\step[fieldset=keywords, fieldvalue={, mypublications}, append]
			}
			\map{
			\perdatasource{AndreAddon.bib}
			\step[fieldset=keywords, fieldvalue={, mypublications}, append]
			}
		}
	}
	\renewbibmacro*{doi+eprint+url}{%
		\iftoggle{bbx:doi}
			{\color{black!40}\footnotesize\printfield{doi}}
			{}%
		\newunit\newblock
		\iftoggle{bbx:eprint}
			{\usebibmacro{eprint}}
			{}%
		\newunit\newblock
		\iftoggle{bbx:url}
			{\usebibmacro{url+urldate}}
			{}%
	}

}

\ifdefined\VersionWithComments
	\usepackage{draftwatermark}
	\SetWatermarkText{draft}
	\SetWatermarkScale{15}
	\SetWatermarkColor[gray]{0.9}
\fi
\usepackage[svgnames,table]{xcolor}
\definecolor{darkblue}{rgb}{0.0,0.0,0.6}
\definecolor{darkgreen}{rgb}{0, 0.5, 0}
\definecolor{darkpurple}{rgb}{0.7, 0, 0.7}
\definecolor{darkblue}{rgb}{0, 0, 0.7}

\usepackage[
		colorlinks=true,
	\ifdefined \VersionWithComments
		pagebackref=true,
	\fi
		citecolor=darkgreen,
		linkcolor=darkblue,
		urlcolor=darkpurple,
	]{hyperref}

\usepackage[capitalise,english,nameinlink]{cleveref} %
\crefname{line}{\text{line}}{\text{lines}} %

\usepackage{array}
\usepackage{tabularx,booktabs}
\newcolumntype{M}[1]{>{\centering\arraybackslash}m{#1}}

\usepackage{listings}
\usepackage{color}

\definecolor{mygreen}{rgb}{0,0.6,0}
\definecolor{mygray}{rgb}{0.5,0.5,0.5}
\definecolor{mymauve}{rgb}{0.58,0,0.82}

\lstdefinestyle{galileo}{
	backgroundcolor=\color{white},   %
	basicstyle=\footnotesize,        %
	breakatwhitespace=false,         %
	breaklines=true,                 %
	captionpos=b,                    %
	commentstyle=\color{mygreen},    %
	deletekeywords={...},            %
	escapeinside={\%*}{*)},          %
	extendedchars=true,              %
	frame=single,	                   %
	keepspaces=true,                 %
	keywordstyle=\color{red!70!black}\bfseries,       %
	morekeywords={and, cost, maxtime, mintime, or, toplevel},            %
	numbers=left,                    %
	numbersep=5pt,                   %
	numberstyle=\tiny\color{mygray}, %
	rulecolor=\color{black},         %
	showspaces=false,                %
	showstringspaces=false,          %
	showtabs=false,                  %
	stepnumber=1,                    %
	stringstyle=\color{mymauve},     %
	tabsize=2,	                   %
}

\usepackage{tikz}
\usetikzlibrary{arrows,automata, arrows.meta, shapes.geometric}
\tikzstyle{every node}=[initial text=]
\tikzstyle{location}=[rectangle, rounded corners, minimum size=12pt, draw=black, fill=blue!10, inner sep=2pt]
\tikzstyle{success}=[fill=green!50]
\tikzstyle{failure}=[fill=red!50]
\tikzstyle{urgent}=[fill=yellow, densely dotted]
\tikzstyle{final}=[double]
\tikzstyle{invariant}=[draw=black, dotted, inner sep=1pt] %
\input{dft-tikz}

\usepackage{wrapfig}

\definecolor{coloract}{rgb}{0.50, 0.70, 0.30}
\definecolor{colorclock}{rgb}{0.4, 0.4, 1}
\definecolor{colordisc}{rgb}{1, 0, 1}
\definecolor{colorloc}{rgb}{0.2, 0.2, 0.35}
\definecolor{colorparam}{rgb}{1, 0.6, 0.0}
\definecolor{colorcost}{rgb}{1, 0.7, 0.3}

\newcommand{\styleact}[1]{\ensuremath{\textcolor{coloract}{\mathrm{#1}}}}
\newcommand{\styleclock}[1]{\ensuremath{\textcolor{colorclock}{\mathrm{#1}}}}

\newcommand{\styleparam}[1]{\ensuremath{\textcolor{colorparam}{\mathrm{#1}}}}

\usepackage{amsthm}
\theoremstyle{plain}
\theoremstyle{definition}
\newtheorem{definition}{Definition}
\newtheorem{example}{Example}

\theoremstyle{remark}
\newtheorem{remark}{Remark}

\usepackage{marginnote}
\ifdefined \VersionWithComments
	\newcommand{\marginX}{\marginnote{\huge{\quad\quad\textbf{!}\quad\quad}}}
	\newcommand{\ea}[1]{\mbox{}{\color{blue}\marginX{}\textbf{[\'Etienne}: #1]}}
	\newcommand{\dl}[1]{\mbox{}{\color{purple}\marginX{}\textbf{[Didier}: #1]}}
	\newcommand{\mr}[1]{\mbox{}{\color{orange}\marginX{}\textbf{[mathias}: #1]}}
	\newcommand{\instructions}[1]{{\color{red}\marginX{}\textbf{[Instructions: ``#1'']}}}
	\newcommand{\reviewer}[2]{\mbox{}{\color{red}\marginX{}\textbf{[Reviewer #1}: ``#2'']}}
	\newcommand{\todo}[1]{\mbox{}{\color{red}{\marginX{}\textbf{TODO}\ifx#1\\\else:\ \fi #1}}} %
\else
	\newcommand{\instructions}[1]{}
	\newcommand{\ea}[1]{}
	\newcommand{\dl}[1]{}
	\newcommand{\mr}[1]{}
	\newcommand{\reviewer}[2]{}
	\newcommand{\todo}[1]{}
\fi

\ifdefined\VersionWithComments
	\usepackage[pagewise, switch]{lineno} %
	\linenumbers
	
\fi

\newcommand{\init}{_0}

\newcommand{\A}{\ensuremath{\mathcal{A}}}
\newcommand{\Actions}{\Sigma}
\newcommand{\action}{\ensuremath{a}}
\newcommand{\aft}[1]{{\ttfamily #1}}

\newcommand{\C}{C}

\newcommand{\Clock}{\mathbb{C}} %
\newcommand{\ClockCard}{H} %
\newcommand{\clock}{x} %
\newcommand{\clockval}{\nu} %
\newcommand{\ClocksZero}{\vec{0}_\Clock}
\newcommand{\compOp}{\bowtie}

\newcommand{\edge}{e}
\newcommand{\Edges}{E}
\newcommand{\longuefleche}[1]{\stackrel{#1}{\longrightarrow}}
\newcommand{\longueflecheRel}[1]{\stackrel{#1}{\mapsto}}

\newcommand{\flecheRel}{{\rightarrow}}

\newcommand{\grandn}{{\mathbb N}}
\newcommand{\grandq}{{\mathbb Q}}
\newcommand{\grandqplus}{\grandq_{+}} %
\newcommand{\grandr}{{\mathbb R}}
\newcommand{\grandrplus}{\grandr_{+}} %

\newcommand{\guard}{g}
\newcommand{\invariant}{I}
\newcommand{\K}{K}

\newcommand{\ArExp}{\mathcal{LA}} %
\newcommand{\loc}{l} %
\newcommand{\locinit}{\loc\init}
\newcommand{\Loc}{L} %
\newcommand{\LocFinal}{F}

\newcommand{\resets}{R}
\newcommand{\TParam}{\mathbb{TP}} %
\newcommand{\tparam}{p} %
\newcommand{\TParamCard}{J} %

\newcommand{\tpval}{tv} %

\newcommand{\weight}{w} %
\newcommand{\weightval}{\mu} %
\newcommand{\Weights}{\mathbb{W}} %
\newcommand{\WeightsCard}{M} %
\newcommand{\WeightsZero}{\vec{0}_\Weights}
\newcommand{\WAssign}{\alpha}
\newcommand{\WValuate}[3]{\ensuremath{\mathit{eval}_{#2}(#1, #3)}} %
\newcommand{\WParam}{\mathbb{WP}} %
\newcommand{\wparam}{q}
\newcommand{\WParamCard}{N} %
\newcommand{\wpval}{wv} %

\newcommand{\partfun}{\nrightarrow}

\newcommand{\sinit}{s\init} %
\renewcommand{\S}{S}
\newcommand{\state}{\ensuremath{s}} %
\newcommand{\States}{S} %
\newcommand{\reset}[2]{\ensuremath{[#1]_{#2}}}
\newcommand{\valuate}[2]{\ensuremath{#2(#1)}}
\newcommand{\twpval}[2]{#1|#2} %
\newcommand{\imitator}{\textsf{IMITATOR}}
\newcommand{\uppaalsmc}{\textsf{Uppaal SMC}}
\newcommand{\galileo}{\textsf{Galileo}}
\newcommand{\uppaal}{\textsf{Uppaal}}

\newcommand{\EFsynth}{\textsf{EFsynth}}
\newcommand{\andgate}{\textsf{AND}}
\newcommand{\orgate}{\textsf{OR}}
\newcommand{\sandgate}{\textsf{SAND}}
\newcommand{\pandgate}{\textsf{PAND}}
\newcommand{\sorgate}{\textsf{SOR}}
\newcommand{\xorgate}{\textsf{XOR}}
\newcommand{\fdepgate}{\textsf{FDEP}}
\newcommand{\sparegate}{\textsf{SPARE}}
\newcommand{\votgate}{\textsf{VOT($k/n$)}}

\newcommand{\pluseq}{\mathrel{+}=}

\ifdefined \VersionWithComments
 	\definecolor{colorok}{RGB}{80,80,150}
\else
	\definecolor{colorok}{RGB}{0,0,0}
\fi

\newcommand{\eg}{\textcolor{colorok}{e.\,g.,}\xspace}

\newcommand{\ie}{\textcolor{colorok}{i.\,e.,}\xspace}
\newcommand{\st}{\textcolor{colorok}{s.t.}\xspace}

\title{Parametric analyses of attack-fault trees%
\thanks{\LongVersion{%
	This manuscript is the extended version of the manuscript of the same name published in the proceedings of the 19th International Conference on Application of Concurrency to System Design (ACSD 2019).
	The final version is available at \url{https://ieeexplore.ieee.org/}.
	}%
	This work is partially supported by the ANR national research program PACS (ANR-14-CE28-0002),
	the PHC Van Gogh project PAMPAS,
	by STW under the project 15474 SEQUOIA,
	KIA KIEM project 628.010.006 StepUp,
	the EU under the project 102112 SUCCESS
	and
	ERATO HASUO Metamathematics for Systems Design Project (No.\ JPMJER1603), JST.
	}
}

\author{\IEEEauthorblockN{\'Etienne Andr\'e}
\IEEEauthorblockA{\textit{Universit\'e Paris 13, LIPN, CNRS, F-93430, Villetaneuse, France}\\
\textit{JFLI, CNRS}, %
Tokyo, Japan %
				/
\textit{National Institute of Informatics}, %
Tokyo, Japan
}
\and
\IEEEauthorblockN{Didier Lime}
\IEEEauthorblockA{\textit{\'Ecole Centrale de Nantes, LS2N, CNRS,} \\
\textit{UMR 6004,} %
Nantes, France
}
\and
\IEEEauthorblockN{Mathias Ramparison}
\IEEEauthorblockA{\textit{Universit\'e Paris 13, LIPN, CNRS,} \\
\textit{UMR 7030, F-93430} %
Villetaneuse, France
}
\and
\IEEEauthorblockN{Mari\"{e}lle Stoelinga}
\IEEEauthorblockA{\textit{Formal Methods and Tools, University of Twente},}
The Netherlands
}

\begin{document}

\ifdefined\VersionLong
	\pagestyle{plain}
\fi

\maketitle

\ifdefined\VersionLong
	\thispagestyle{plain}
\fi

\ifdefined \VersionWithComments
	\textcolor{red}{\textbf{This is the version with comments. To disable comments, comment out line~3 in the \LaTeX{} source.}}
\fi

\begin{abstract}
Risk assessment of cyber-physical systems, such as power plants, connected devices and IT-infrastructures has always been challenging: safety (\ie{} absence of unintentional failures) and security (\ie{} no disruptions due to attackers) are conditions that must be guaranteed.
	One of the traditional tools used to help considering these problems is attack trees, a tree-based formalism inspired by fault trees, a well-known formalism used in safety engineering.
	In this paper we define and implement the translation of attack-fault trees (AFTs) %
		to a new extension of timed automata, called parametric weighted timed automata.
	This allows us to parametrize constants such as time and discrete costs in an AFT and then, using the model-checker \imitator{}, to compute the set of parameter values such that a successful attack is possible.
	Using the different sets of parameter values computed, different attack and fault scenarios can be deduced depending on the budget, time or computation power of the attacker, providing helpful data to select the most efficient counter-measure.
\end{abstract}


\section{Introduction}\label{section:introduction}

\todo{TODO mathias : voir pour remplacer attacker profiles par attack scenarios partout, ça semble plus pertinent puisqu'on peut faire aussi bien du attack tree que du fault tree}

In the past few years, the range of security breaches in the security of organizations has become larger and larger. The process of unifying them by determining relations and consequences between separated events has become more difficult: how to relate
the presence of solid oxygen in a helium tank in SpaceX rocket Falcon 9 to its the explosion during firing tests?
What is the cost for the attacker and the damages caused to SpaceX manufacturing plants?
	\dl{tu as quelque chose en tête de plus précis, notamment pour "special device"? là j'ai un peu du mal à comprendre cet exemple}\ea{idem; si tu penses à quelque chose de réel, tu peux mettre une ref (si scientifique) ou une footnote (article sur Internet)}\ea{et donc?}\mr{en fait aucune idée}%
One of the tools available to help structure risk assessments and security analyses is \emph{attack trees}, recommended, \eg{} by NATO Research and Technology Organisation (RTO) \cite{NATO} and OWASP (Open Web Application Security Project)~\cite{OWASP}.
\emph{Attack trees} \cite{SSSW98} were formalized in~\cite{KMRS10} as a popular and convenient formalism for security analysis (see \cite{KPS14} for a survey) and are inspired by \emph{fault trees} \cite{FMC09,RS15} a well-known formalism used in safety engineering.
Bottom-up computation for a single parameter (\eg{} cost, probability or time of an attack), can be performed directly on attack trees \cite{BKMS12}.
Attack trees and fault trees are quite similar but differ on their gates and/or goals \cite{BKMS12,KMRS14}.
Both are constructed with leaves that model component and attack step failures or successes that propagate through the system via gates.
While fault trees focus on safety properties, attack trees considerate skills, resources and risk appetite possessed by an attacker performing actions.
\emph{Attack-fault trees} (AFTs) \cite{KS17} combine safety properties from fault trees and security conditions from attack trees; therefore gates of both fault trees and attack trees are used in this formalism.

Quantitative analysis of AFTs with multiple quantitative annotations on AFTs like cost, time, failure probabilities---which can functionally be dependent on each other---evaluates risks and helps to determine the most risky scenarios and therefore to select the most effective counter-measures.

\paragraph{Contribution}
In this work, we study a more abstract version of the security problem, and we propose an approach to \emph{synthesize} times and costs necessary to individual actions in order to perform a successful attack or individual failures causing the failure of the entire system.
The global attack time and cost can then be expressed as a combination of the parametric unit costs.
To this end, we propose a formalization of attack-fault trees using an \emph{ad-hoc} extension of parametric timed automata called \emph{parametric weighted timed automata} (PWTAs).
PWTAs can be seen as a generalization of parametric timed automata (PTAs)~\cite{AHV93} and weighted/priced automata \cite{BFHLPRV01,ATP04} with only costs on transitions.

We implement our framework within the tool ATTop presented in%
	~\cite{SRYHBRS18}, allowing to define AFTs in the \galileo{} format, and provide an automated translation into the \imitator{} input format~\cite{AFKS12}.

As a proof of concept, we apply our framework to an attack tree of \cite{KS17} and an original attack-fault tree. With the help of the parametric timed model checker \imitator{}, we are able to synthesize constraints in several dimensions; further we discuss induced possible attack and fault scenarios.

This enlarges the scope of quantitative analysis for AFTs by parameterizing multiple annotations on the AFT \emph{at once} such as time, cost and damages and then compute for instance the optimal combination of parameter values for the attack to fail quickly while keeping damages to the system low.

\paragraph{Related work}
Attack tree analysis has been studied through lattice theory \cite{KMRS10}, timed automata \cite{KRS15,KS17,SRYHBRS18}, I/O-IMCs \cite{KGS15,AGKS15}, Bayesian networks \cite{GIM15}, Petri nets \cite{DMCR06}, stochastic games \cite{ANP16,HKKS16}, etc.
\uppaal{} has been used for model transformations in \cite{SYRGKDRS17} and in \cite{HV06} UML sequence diagrams are manually transformed into timed automata models.
\cite{KS17} especially tackles the problem of multiple complex risk metrics and attacker profiles, in a probabilistic and timed formalism that can be computed and analyzed using stochastical model-checking \cite{RS12}\ea{c'est bien tous les actes de l'école que tu veux citer ?}\mr{c'est bien la ref [20] de \cite{KS17} il me semble} and \uppaalsmc{} \cite{DLLMP15}.\dl{ça ne va pas dans les related work ça?}\ea{ça m'est égal ; si tu déplaces, Mathias, il faut déplacer la phrase qui suit avec}
AFTs are modeled in the \galileo{} format and translated with the tool ATTop \cite{SRYHBRS18} into stochastic timed automata \cite{DLLMPVW11}.

However, synthesis of multidimensional parameters (time, cost for the attacker, damages for the organization...) at once for fully timed systems is not treated in the previously cited works, and these works require testing one by one a set of possible attribute values for an AFT.

Besides, attack-defense trees are one of the most well-studied extensions of attack trees and new analysis methods are still developed~\cite{KMRS14,KW18}.

In a completely different area, asynchronous hardware circuits' gates were translated into (parametric) timed automata in~\cite{CEFX09}; our translation of AFTs gates into PWTAs synchronized using parallel composition shares some similarities with that approach.\ea{à virer si pas de place}

\paragraph{Outline}
We recall attack-fault trees in \cref{section:AFT}.
We then introduce the formalism of \emph{parametric weighted timed automata} in \cref{section:PWTA}.
Our translation from AFTs to PWTAs is given in \cref{section:translation}.
Then, we describe our implementation in \cref{section:implementation} and report on experiments in \cref{section:casestudies}.
We conclude by discussing future works.\ea{ ``in \\cref\{section:finalremarks\}'' là y'a un bug de fou, si je mets la bonne réf de section (avec cref ou ref) ça fait tout planter ; si je mets une broken ref ou pas de ref, pas de bug… Je comprends rien, et j'abandonne} %
\dl{chez moi ça passe...}

\section{Attack-fault Trees}\label{section:AFT}

Attack-fault trees (AFTs) model how a safety or security goal can be refined into smaller sub-goals, represented as \emph{gates}, until no further refinement is possible, represented as \emph{leaves}.
The leaves of the tree model are either basic component failures (BCF) or basic attack steps (BAS).
Since subtrees can be shared in the literature (see \eg{} \cite{KS17}), AFTs are actually directed acyclic graphs, rather than trees. In this paper, we consider only trees without shared gates or leaves.
Safety is compromised with the failure of a BCF, \ie{} without any outside spark action.
Security is compromised when an outside attacker causes the activation of a BAS.
Following the terminology of \cite{KS17}, in this paper write that a gate or a leaf is \emph{disrupted} if the output is true \ie{} it succeeds, and \emph{fails} otherwise. A success event (disruption) models the fact that a component (gate or leaf) is compromised \ie{} the attack is successful or the component fails. In contrast, a fail event models the robustness of the component against an attacker through a BAS, or a BCF.

\todo{ici il faut bien expliquer la différence entre disruption (= success de l'attaque) et failure (= échec de l'attaque)}\mr{je l'ai fait pour la \andgate{} en dessous, il vaut mieux le mettre dès le début ?}
\dl{je pense que la phrase du todo, ici, ce serait bien}

\subsection{AFT leaves}

AFT leaves are equipped with an execution time and a rich cost\ea{je réalise que tu utilises cost alors que j'utilise weight :p Ça peut aller si chacun des termes s'applique dans son propre domaine (cost pour AFT et weight pour PWTA); mais il faudra bien faire attention à la cohérence} structure that includes the cost incurred by an attacker and the damage inflicted on the organization.
In contrast to~\cite{RS15, KS17} where BCF and BAS are equipped with probability distributions, we consider both BCF and BAS as parametric time-dependent events.
This allows us to compute a range of cost values, damages values and time intervals at once in order to perform operations such as optimum time values for a counter-measure while keeping damage to the organization low, and cost for the attacker high.

\subsection{AFT gates}

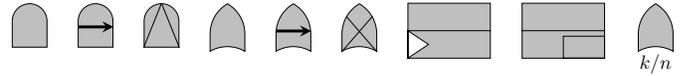
\begin{figure}
\definecolor{attackedcolor}{gray}{0.75}
  \resizebox{\linewidth}{!}{%
\begin{tikzpicture}[
node distance=12mm,
rk/.style={rectangle,draw,minimum
height=5mm,fill=black!10, outer sep=0pt,
font=\ttfamily},
zigzag/.style={to path={|- ($(\tikztostart)!.75!(\tikztotarget)$) -| (\tikztotarget)}}]
\node (AND) {};
\path(AND.south) node[anchor=east,and1, fill=attackedcolor] (G5) {};
\node[right of=AND] (SAND) {};
\path(SAND.south) node[anchor=east,and1, fill=attackedcolor] (G6) {};
\path[->,line width=1.5pt]
(G6.north) edge (G6.south);
\node[right of=SAND] (PAND) {};
\path(PAND.south) node[anchor=east,and1, fill=attackedcolor] (G7) {};
\path[-,line width=.5pt]
(G7.south west) edge (G7.east)
(G7.north west) edge (G7.east);
\node[right of=PAND] (OR) {};
\path(OR.south) node[anchor=east,or1, fill=attackedcolor] (G8) {};
\node[right of=OR] (SOR) {};
\path(SOR.south) node[anchor=east,or1, fill=attackedcolor] (G9) {};
\path[->,line width=1.5pt]
(G9.north) edge (G9.south);
\node[right of=SOR] (XOR) {};
\path(XOR.south) node[anchor=east,or1, fill=attackedcolor] (G10) {};
\path[-,line width=.5pt]
(G10.south west) edge (G10.north east)
(G10.north west) edge (G10.south east);
\node[right of=XOR,xshift=-9] (FDEP) {};
\path(FDEP.south) node[anchor=north west,fdep, fill=attackedcolor] (G11) {};
\node[right of=FDEP, xshift=25] (SPARE) {};
\path(SPARE.south) node[anchor=north west,spare, fill=attackedcolor] (G12) {};
\node[right of=SPARE, xshift=35,label={[align=left, yshift=-23]below:{$k/n$}}] (VOT) {};
\path(VOT.south) node[anchor=east,or1, fill=attackedcolor] (G13) {};
\end{tikzpicture}%
}
\caption{From left to right: \andgate{}, \sandgate{}, \pandgate{}, \orgate{}, \sorgate{}, \xorgate{}, \fdepgate{}, \sparegate{}, \votgate{} gates}
\label{gates}
\end{figure}

In order to model complex scenarios with multiple leaves, BCF and BAS have to be composed.
For this purpose, logical gates are used that output either the propagation of a disruption, or not.
Gates take as an input either leaves or outputs from gates in their subtrees.
Logical gates used in AFTs are taken from both dynamic fault trees and attack trees: \andgate{}, \pandgate{}, \sandgate{}, \orgate{}, \sorgate{}, \fdepgate{}, \sparegate{}, \votgate{}, depicted in \cref{gates}. These gates are the translatable ones in ATTop \cite{SRYHBRS18} from the \galileo{} format. We also added the \xorgate{} gate to improve our modeling capabilities.\ea{si c'est exactement l'ensemble de portes considéré dans un papier existant, je pense qu'il faut le citer ici}\mr{ce sont les portes traductibles depuis un fichier \galileo{} dans ATTop donc sûrement \cite{SRYHBRS18} mais à confirmer}
\mr{en fait il y a la porte XOR que j'ai faite mais qui n'est pas dans ATTOP, je laisse ?}\ea{je dirais oui, mais tu peux aussi préciser ``to which we added the \xorgate{} to improve the modeling capabilities''}

    \andgate{} gate propagates a disruption (\ie{}, it synchronizes a success event \cite{KS17}) if all of its children are disrupted, regardless of the order of disruption.
		Children are activated initially by the \andgate{}.
    Children of a \sandgate{} gate are activated sequentially from left to right. After the success (disruption)\dl{on en remet une couche c'est ça la pédagogie!} of the leftmost child, the second left most child is activated, and so on until the disruption of rightmost child. If \emph{all} children are disrupted, the \sandgate{} gate is disrupted. However, if any child fails (to be disrupted)\dl{là itou}, the \sandgate{} gate directly fails.\mr{dans \cite{KS17} ils utilisent disruption pour dire que la porte/BAS est activée donc success. J'ai utilisé termination pour dire que c'est fini, peu importe si ça fail/success}\ea{mais je suppose que, si un enfant n'est pas disrupté, alors son ``frère cadet'' ne sera pas activé, ou bien ? En tout cas, ça demande plus d'explicaion que pour un bête AND}\ea{Merci des explications. Mais du coup je comprends vraiment pas : quelle est la différence entre disruptée et failed ? et success ? Je crois qu'il faut repréciser (c'est ptêt le cas mais si oui on peut encore insister) qu'un attack tree est successful si l'attaque est successful, et une porte fails si l'attaque fails.}
    \sandgate{} gate is a specific gate of attack trees.
    Compared with \sandgate{} gate, all children of a \pandgate{} gate are activated initially when the \pandgate{} gate is activated\ea{system start ou activation de la porte ?}\dl{et du coup quelle différence avec \andgate{}?}\mr{pas besoin que ce soit de gauche à droite ! c'est la phrase suivante}. The rest of the execution is similar to a \sandgate{} gate, and propagates a disruption if all children are disrupted from left to right (which in contrast is not mandatory for an \andgate{} gate\mr{je rajoute ça ?}), otherwise the \pandgate{} gate fails.

\orgate{} gate propagates a disruption if at least one of its children is disrupted. Children are activated initially by the \orgate{} gate.
    Similarly to a \sandgate{} gate, children of a \sorgate{} gate are activated sequentially after the termination of the previous one and from left to right. It propagates a disruption \emph{when} one of its children is disrupted, otherwise if all children fail the \sorgate{} gate fails.
    \xorgate{} gate propagates a disruption if one of its children is disrupted and the other one fails.

\fdepgate{} (functional dependency) gate consists of a trigger event and several dependent events, and is a specific gate of fault trees. When the trigger event occurs, all its dependent BCF events are disrupted (\ie{} the failure of the power supply automatically deactivate the alarm and security cameras, therefore the BCFs are successful)\dl{là aussi c'est fail to be disrupted ou c'est fail au sens safety?}.

\sparegate{} gate is similar to \sandgate{}, but is a specific gate for fault events while \sandgate{} gate is used for attack events.
\sparegate{} gate consists of one primary BCF and several secondary BCF which are activated sequentially. If the primary BCF is disrupted (\ie{} the component fails), a secondary becomes primary. If no BCFs are left (they all are disrupted), \sparegate{} gate propagates a disruption.\dl{là aussi j'ai envie de comprendre les fail comme de la safety. Je me trompe? Si non alors ce serait plutôt are disrupted}
\mr{exact}
\dl{tu ne disais pas plutôt \sorgate{} dans ta remarque plus haut??}\mr{j'avais mal compris je pense. d'ailleurs comme ils sont activés sequentiellement c'est un \sandgate{}. Au final c'est même exactement un \sandgate{}}

\votgate{} gate is similar to \orgate{} gate and consists of~$n\in\grandn$ children initially activated. \votgate{} gate is disrupted when~$k$ of its~$n$ children are disrupted.\ea{là, ils sont tous activés dès le début ?}

\begin{figure*}[t]
\definecolor{attackedcolor}{gray}{0.75}
  \resizebox{\linewidth}{!}{%
\begin{tikzpicture}[
node distance=12mm,
rk/.style={rectangle,draw,minimum
height=5mm,fill=black!10, outer sep=0pt,
font=\ttfamily},
zigzag/.style={to path={|- ($(\tikztostart)!.75!(\tikztotarget)$) -| (\tikztotarget)}}]
\node[rk, fill=attackedcolor] (Top) {compromise\_IoT\_device};
\path(Top.south) node[anchor=east,and1, fill=attackedcolor] (G5) {};
\node[below of=G5, yshift=1mm, rk, fill=attackedcolor] (SISF) {exploit\_software\_vulnerability\_in\_IoT\_device};
\node[anchor=north east, outer sep=0pt, yshift=-3pt] at (SISF.south east)
{cost = 60 US\$, duration = 1 hour};
\node[anchor=east, rk,xshift=-5mm, fill=attackedcolor,
minimum width=34mm] at
(SISF.west) (PL) {access\_home\_network};
\node[anchor=west ,rk,xshift=5mm, fill=attackedcolor] at
(SISF.east) (SI) {run\_malicious\_script};
\node[anchor=north east] at (SI.south east)
{\begin{tabular}{r}cost = 50
US\$\\ duration = 0.5 hour\end{tabular}};
\path (PL.south) node[anchor=east,and1, fill=attackedcolor] (G15) {};
\path (G15.input 2 -| G5.input 2) ++ (0,-6mm) node[rk, fill=attackedcolor] (SFL) {gain\_access\_to\_private\_networks};
\path (G15.west) ++ (0, -6mm) node[rk,minimum width=34mm, label={[align=left]below:{cost = 40 US\$,\\ duration = 10 hours}}, fill=attackedcolor] (SFF) {get\_credentials};
\path (SFL.south) node[anchor=east,or1, fill=attackedcolor] (G16) {};
\node[below of=G16, rk,xshift=-34mm, yshift=4mm] (SVL) {access\_LAN};
\node[below of=G16, rk,xshift=39mm, yshift=4mm, fill=attackedcolor] (SWL) {access\_WLAN};
\path (SVL.south) node[anchor=east,and1] (G17) {};
\node[minimum width=36mm, yshift=2mm, below of=G17, rk,xshift=-2mm, anchor=east,label={[align=left]below:{cost = 20 US\$,\\ duration = 1 hour}}] (SLP) {find\_LAN\_access\_port};
\node[minimum width=36mm, yshift=2mm, below of=G17, rk,xshift=2mm, anchor=west,label={[align=left]below:{cost = 30 US\$,\\ duration = 0.5 hour}}] (SSP) {spoof\_MAC\_address};
\path (SWL.south) node[anchor=east,and1, fill=attackedcolor] (G18) {};
\node[minimum width=30mm, yshift=2mm, below of=G18, rk,xshift=-2mm, anchor=east,label={[align=left]below:{cost = 2 US\$,\\ duration = 5 hours}}, fill=attackedcolor] (SP) {find\_WLAN};
\node[minimum width=30mm, yshift=2mm, below of=G18, rk, anchor=west,label={[align=left]below:{cost = 80 US\$,\\ duration = 2 hours}}, fill=attackedcolor] (SS) {break\_WPA\_keys};

\node[font=\large, anchor=south east] at (SS.north east)
{\color{red}0--2};
\node[font=\large, anchor=south west] at (SP.north west)
{\color{red}0--5};
\node[font=\large, anchor=south west] at (SFF.north west)
{\color{red}0--10};
\node[font=\large, anchor=north west] at (SISF.south west)
{\color{red}10--11};
\node[font=\large, anchor=south east] at (SI.north east)
{\color{red}11--11.5};

\path[draw] (G5.west) -- ++(0,-4mm) -- (SISF.north -| G5.west);
\path[-]
(PL) edge[zigzag] (G5.input 1)
(SI) edge[zigzag] (G5.input 2)
(SFF.north -| G15.input 1) edge[zigzag] (G15.input 1)
(SFL) edge[zigzag] (G15.input 2)
(SVL) edge[zigzag] (G16.input 1)
(SWL) edge[zigzag] (G16.input 2)
(SLP) edge[zigzag] (G17.input 1)
(SSP) edge[zigzag] (G17.input 2)
(SP) edge[zigzag] (G18.input 1)
(SS) edge[zigzag] (G18.input 2)
;
\path[->,line width=1.5pt]
(G5.north) edge (G5.south)
;
\end{tikzpicture}%
}
\caption{Attack Tree modeling the compromise of an IoT device from \cite{SYRGKDRS17}.
Leaves are equipped with the cost and time required to execute
the corresponding step.
The parts of the tree attacked in a
successful attack are indicated by a darker color, with
start and end times for the steps in this attack denoted in
{\color{red}red}.}
\label{ATree}
\end{figure*}
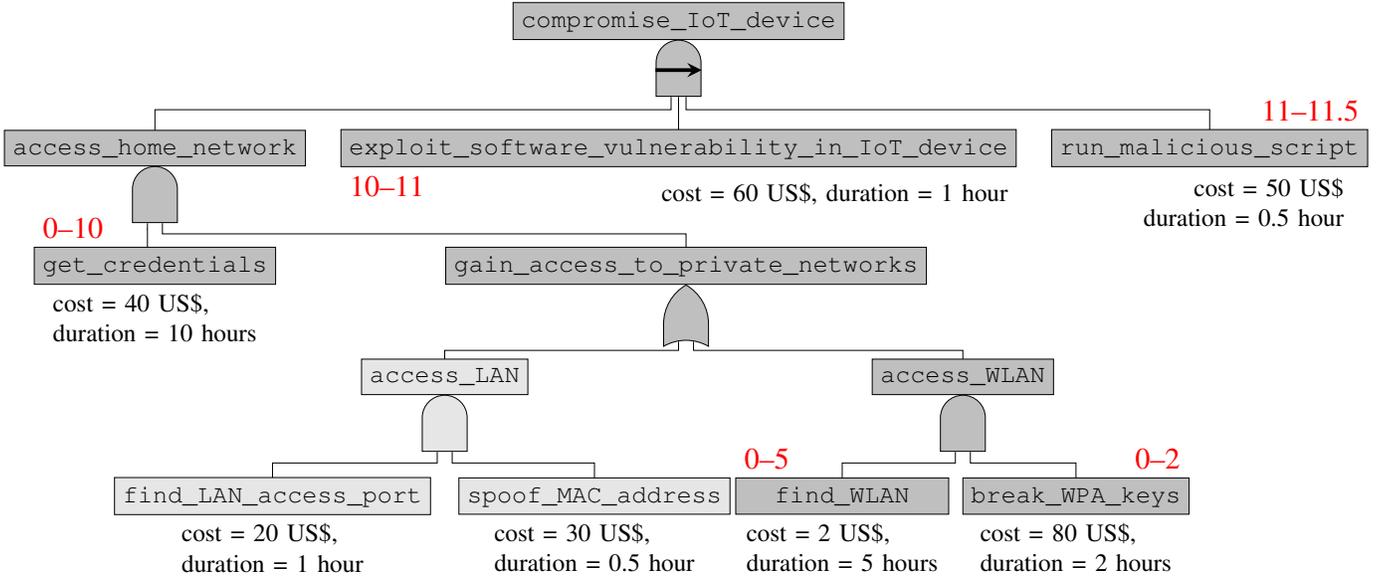

\section{Parametric weighted timed automata}\label{section:PWTA}

Let $\grandn$, $\grandq$, $\grandqplus$,  and $\grandrplus$ denote the set of non-negative integers, rationals, non-negative rationals and non-negative reals, respectively.

We assume a set~$\Clock = \{ \clock_1, \dots, \clock_\ClockCard \} $ of \emph{clocks}, \ie{} real-valued variables that evolve at the same rate.
A \emph{clock valuation} is a function
$\clockval : \Clock \rightarrow \grandrplus$.
We write $\ClocksZero$ for the clock valuation assigning $0$ to all clocks.
Given $d \in \grandrplus$, $\clockval + d$ \ShortVersion{is}\LongVersion{denotes the valuation} \st{} $(\clockval + d)(\clock) = \clockval(\clock) + d$, for all $\clock \in \Clock$.
Given $\resets \subseteq \Clock$, we define the \emph{reset} of a valuation~$\clockval$, denoted by $\reset{\clockval}{\resets}$, as follows: $\reset{\clockval}{\resets}(\clock) = 0$ if $\clock \in \resets$, and $\reset{\clockval}{\resets}(\clock)=\clockval(\clock)$ otherwise.

We assume a set~$\TParam = \{ \tparam_1, \dots, \tparam_\TParamCard \} $ of \emph{timing parameters}\LongVersion{, \ie{} unknown timing constants}.
A \emph{timing parameter valuation} $\tpval$ is\LongVersion{ a function}
$\tpval : \TParam \rightarrow \grandqplus$.
We assume ${\compOp} \in \{<, \leq, =, \geq, >\}$.
A \emph{guard}~$\guard$ is a constraint over $\Clock \cup \TParam$ defined by a conjunction of inequalities of the form $\clock \compOp d$,
or $\clock \compOp \tparam$ with $\clock \in \Clock$, $d \in \grandn$ and $\tparam \in \TParam$.
Given~$\guard$, we write~$\clockval\models\tpval(\guard)$ if %
the expression obtained by replacing each~$\clock$ with~$\clockval(\clock)$ and each~$\tparam$ with~$\tpval(\tparam)$ in~$\guard$ evaluates to true.

We assume a set~$\Weights = \{ \weight_1, \dots, \weight_\WeightsCard \} $ of \emph{weights}.
A \emph{weight valuation} $\weightval$ is\LongVersion{ a function}
$\weightval : \Weights \rightarrow \grandq$.
We write $\WeightsZero$ for the weight valuation assigning $0$ to all weights.
We assume a set~$\WParam = \{ \wparam_1, \dots, \wparam_\WParamCard \} $ of \emph{weight parameters}\LongVersion{, \ie{} unknown weight constants}.
A \emph{weight parameter valuation} $\wpval$ is\LongVersion{ a function}
$\wpval : \WParam \rightarrow \grandq$.\footnote{%
    Observe that, in contrast to timing parameters that should be non-negative (which is usual for \emph{parametric timed automata}), our weight parameters may be negative.\mr{on l'utilise où ça ?}
}
A linear arithmetic expression over $\Weights \cup \WParam$ is $\sum_i a_i \weight_i + \sum_j b_j \wparam_j + c$, where $\weight_i \in \Weights$, $\wparam_j \in \WParam$ and $a_i, b_j, c \in \grandq$.
Let $\ArExp(\Weights \cup \WParam)$ denote the set of arithmetic expressions over $\Weights$ and~$\WParam$.
A parametric weight update is a partial function
	$\WAssign : \Weights \partfun \ArExp(\Weights \cup \WParam)$.
	That is, we can assign a weight to an arithmetic expression of parametric weights and other weight values, and rational constants.
Given a weight valuation~$\weightval$, a parametric weight update~$\WAssign$ and a weight parameter valuation~$\wpval$,
	we need an evaluation function $\WValuate{\WAssign}{\wpval}{\weightval}$ returning a weight valuation, and defined as follows:
\[\WValuate{\WAssign}{\wpval}{\weightval}(\weight) =
	\begin{cases}
		\weightval(\weight) & \text{if $\WAssign(\weight) $ is undefined} \\
		\valuate{\valuate{\WAssign(\weight)}{\wpval}}{\weightval} & \text{otherwise} \\
	\end{cases}
\]
where $\valuate{\valuate{\WAssign(\weight)}{\wpval}}{\weightval}$ denotes the replacement within the linear arithmetic expression $\WAssign(\weight)$ of all occurrences of a weight parameter~$\wparam_i$ by~$\wpval(\wparam_i)$, and of a weight variable $\weight_j$ with its current value $\weightval(\weight_j)$.
Observe that this replacement gives a rational constant, therefore $\WValuate{\WAssign}{\wpval}{\weightval}$ is indeed a weight valuation $\Weights \rightarrow \grandq$.
That is, $\WValuate{\WAssign}{\wpval}{\weightval}$ computes the new (non-parametric) weight valuation obtained after applying to~$\weightval$ the partial function $\WAssign$ valuated with~$\wpval$.

Parametric timed automata (PTA) extend timed automata~\cite{AD94} with timing parameters\LongVersion{ within guards and invariants in place of integer constants}~\cite{AHV93}.
We extend further PTA with (discrete) rational-valued \emph{weight parameters}, giving birth to parametric weighted timed automata (PWTA).
\begin{definition}%
	\label{def:PWTA}
	A parametric weighted timed automaton (PWTA) $\A$ is a tuple \mbox{$\A = (\Actions, \Loc, \locinit, \LocFinal, \Clock, \TParam, \Weights, \WParam, \invariant, \Edges)$}\LongVersion{, where}:
	\begin{enumerate}
		\item $\Actions$ is a finite set of synchronization actions,
		\item $\Loc$ is a finite set of locations,
		\LongVersion{\item} $\locinit \in \Loc$ is the initial location,
		\item $\LocFinal \subseteq \Loc$ is the set of accepting locations,
		\item $\Clock$ is a finite set of clocks,
		\item $\TParam$ is a finite set of timing parameters,
		\item $\Weights$ is a finite set of weights,
		\item $\WParam$ is a finite set of weight parameters,
		\item $\invariant$ is the invariant, assigning to every $\loc\in \Loc$ a guard $\invariant(\loc)$,
		\item $\Edges$ is a finite set of edges  $\edge = (\loc, \guard, \action, \resets, \WAssign, \loc')$
		where~$\loc,\loc'\in \Loc$ are the source and target locations,
			$\guard$ is a guard,
			$\action \in \Actions$,
			$\resets\subseteq \Clock$ is a set of clocks to be reset,
			and
			$\WAssign : \Weights \partfun \ArExp(\Weights \cup \WParam)$ is a parametric weight update.
	\end{enumerate}
\end{definition}

Given a timing parameter valuation~$\tpval$ and a weight parameter valuation~$\wpval$, we denote by $\valuate{\A}{\twpval{\tpval}{\wpval}}$ the non-parametric structure where all occurrences of a timing parameter~$\tparam_i$ have been replaced by~$\tpval(\tparam_i)$, and all occurrences of a weight parameter~$\wparam_j$ have been replaced by~$\wpval(\wparam_j)$.
The resulting structure can be seen as an extension of a parametric weighted/priced timed automaton \cite{BFHLPRV01,ATP04} with only rational weights on edges.\footnote{%
	In \cite{ATP04} cost is defined as the sum of each discrete cost on transitions (\emph{switch} cost) plus the time spent in a location multiplied by an integer rate (\emph{duration} cost), resulting in a rational value.
	Here, we omit the duration costs.}
However, our structure goes beyond a simple parametric extensions of weighted/priced timed automata, for two reasons:
\begin{ienumeration}
	\item we allow multiple weights;
	\item we allow to not only \emph{increment} weight values over a path, but also perform more complex operations on that weight, notably incrementing it \emph{with another weight value}, which is clearly not possible in \cite{BFHLPRV01,ATP04}.
\end{ienumeration}
Note that, if we restrict our parametric weight update function to expressions of the form $\WAssign(\weight_i) = \weight_i + z$, where $z$ is either a weight parameter or a rational constant, then our formalism is exactly the parametric extension of (the discrete ``switch'' weight part of) \cite{BFHLPRV01,ATP04}.%
\footnote{%
    Technically, as weighted/priced timed automata\dl{weighted pour Alur et priced pour Larsen donc la formulation est bonne!}  use integer constants, a rescaling of the constants is necessary: by multiplying all constants in $\valuate{\A}{\twpval{\tpval}{\wpval}}$ by the least common multiple of their denominators\dl{bon j'ai mis la bonne formulation ici: c'est le PPCM des dénominateurs :-)}, we obtain an equivalent (integer-valued) weighted/priced timed automata.
}

	In addition, our formalism shares some similarities with the statically parametric timed automata of~\cite{Wang00}, where timed automata are extended with parameters that can only be used in guards, but not compared to clocks.
	In contrast, our weight parameters can only be used in updates, and not in guards; in addition, we also feature the timing parameters of~\cite{AHV93} that can be compared to clocks.

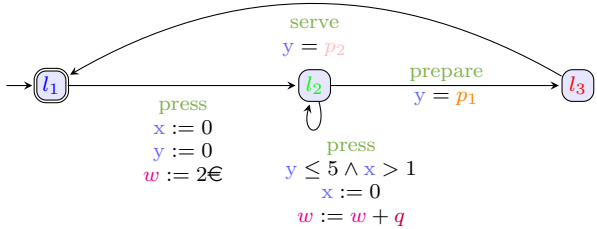
\begin{figure}[h!]
\footnotesize
\begin{center}
\begin{tikzpicture}[shorten >=1pt, node distance=3.5cm, on grid, auto]
   \node[location, initial, final, initial text={$\styleclock{x},\styleclock{y} := 0$}, inner sep=1pt] (A)   {$\color{blue}l_{1}$};
   \node[location]          (B)   [right=of A]           {$\color{green}l_{2}$};
   \node[location]          (C)   [right=of B]           {$\color{red}l_{3}$};
  \path[->]
             (A)  edge   node  [swap]   {\begin{tabular}{c}
  											\styleact{press}\\
											 $\styleclock{x} := 0$\\
											 $\styleclock{y} := 0$\\
											 ${\color{magenta}\weight}:=2$\euro
 											\end{tabular}}   (B)
             (B)  edge   node[above, yshift=-4mm] {\begin{tabular}{c}
  											\styleact{prepare}\\
											 $\styleclock{y}={\color{orange}p_{1}}$
 											\end{tabular}}   (C)
			(B)	edge   [loop below]  node  [right, yshift=-7mm, xshift=-7mm]{\begin{tabular}{c}
  											\styleact{press}\\
											 $\styleclock{y}\leq 5 \land \styleclock{x}>1$\\
											 $\styleclock{x}:=0$\\
											 ${\color{magenta}\weight}:={\color{magenta}\weight}+{\color{purple}\wparam}$
 											\end{tabular}}   ()
             (C)    edge [bend right]  node[below] {\begin{tabular}{c}
  											\styleact{serve}\\
											 $\styleclock{y}={\color{pink}p_{2}}$
 											\end{tabular}}   (A);
\end{tikzpicture}
\end{center}
\caption{A PWTA modeling a coffee machine}
\label{figure:PTAcoffee}
\end{figure}

\begin{example}\label{example:motivating}
	In the PWTA in \cref{figure:PTAcoffee}, we have the following elements: $\Loc=\lbrace {\color{blue}l_{1}}, {\color{green}l_{2}}, {\color{red}l_{3}}\rbrace$, $\locinit= {\color{blue}l_{1}}$ (also the unique element of $\LocFinal$),  $\Clock=\lbrace \styleclock{x} , \styleclock{y} \rbrace$, and $\Actions=\lbrace \styleact{press}, \styleact{prepare}, \styleact{serve}\rbrace$, with the set~$\TParam=\lbrace {\color{orange}p_{1}}, {\color{pink}p_{2}}\rbrace$ and weights $\Weights=\lbrace {\color{magenta}\weight}\rbrace$, $\WParam=\lbrace{\color{purple}\wparam}\rbrace$.
	There are four edges:
	\begin{itemize}
	\item $\edge_1 = \langle{\color{blue}l_{1}},\guard,\action,\resets,{\color{green}l_{2}}\rangle$ where~$\resets$ sets both~$\styleclock{x},\styleclock{y}$ to~$0$, $\WAssign$ is ${\color{magenta}\weight}=2$\euro,
	\item $\edge_2 = \langle{\color{green}l_{2}},\guard,\action,\resets,{\color{green}l_{2}}\rangle$ where~$\guard$ is~$\styleclock{y}\leq 5\wedge \styleclock{x}>1$ and~$\resets$ sets~$\styleclock{x}$ to~$0$, $\WAssign$ is ${\color{magenta}\weight}:={\color{magenta}\weight}+{\color{purple}\wparam}$,
	\item $\edge_3 = \langle{\color{green}l_{2}},\guard,\action,\resets, {\color{red}l_{3}}\rangle$ where~$\guard$ is~$\styleclock{y}={\color{orange}p_{1}}$ and
	\item $\edge_4 = \langle{\color{red}l_{3}},\guard,\action,\resets, {\color{blue}l_{1}}\rangle$ where~$\guard$ is~$\styleclock{y}={\color{pink}p_{2}}$.
	\end{itemize}
\end{example}

Let us now define the concrete semantics of PWTA as the union over all timing parameter and weight parameter valuations.

\begin{definition}[Semantics of a valuated PWTA]
	Given a PWTA $\A = (\Actions, \Loc, \locinit, \LocFinal, \Clock, \TParam, \Weights, \WParam, \invariant, \Edges)$,
	a timing parameter valuation~\(\tpval\),
	and
	a weight parameter valuation~\(\wpval\),
	the semantics of $\valuate{\A}{\twpval{\tpval}{\wpval}}$ is given by the timed transition system (TTS) $(\States, \sinit, \flecheRel)$, with
	\begin{itemize}
		\item $\States = \{ (\loc, \clockval, \weightval) \in \Loc \times \grandrplus^\ClockCard \times \grandq^\WeightsCard \mid \clockval \models \valuate{\invariant(\loc)}{\tpval} \}$, %
		\item $\sinit = (\locinit, \ClocksZero, \WeightsZero) $,
		\item  $\flecheRel$ consists of the discrete and (continuous) delay transition relations:
		\begin{ienumeration}
			\item discrete transitions: $(\loc, \clockval, \weightval) \longueflecheRel{\edge} (\loc',\clockval', \weightval')$, %
				if
					$(\loc, \clockval, \weightval) , (\loc',\clockval', \weightval') \in \States$, and %
				there exists $\edge = (\loc, \guard, \action, \resets, \WAssign, \loc') \in \Edges$, such that
					$\clockval \models \tpval(\guard$),
					$\clockval'= \reset{\clockval}{\resets}$,
				and
					$\weightval' = \WValuate{\WAssign}{\wpval}{\weightval}(\weight)$;
			\item delay transitions: $(\loc, \clockval, \weightval) \longueflecheRel{d} (\loc, \clockval+d, \weightval)$, with $d \in \grandrplus$, if $\forall d' \in [0, d], (\loc, \clockval+d', \weightval) \in \States$.
		\end{ienumeration}
	\end{itemize}
\end{definition}

That is, a state is a triple made of the current location, the current (non-parametric) clock valuation, and the current (non-parametric) weight valuation.
The clock valuations evolve naturally as in timed automata, while the current weight
	evolves according to the weight update function.

    Moreover we write $(\loc, \clockval, \weightval)\longuefleche{(\edge, d)} (\loc',\clockval', \weightval')$ for a combination of a delay and discrete transition if
		$\exists  \clockval'' :  (\loc, \clockval, \weightval) \longueflecheRel{d} (\loc, \clockval'', \weightval) \longueflecheRel{\edge} (\loc', \clockval', \weightval')$.
Given $\valuate{\A}{\twpval{\tpval}{\wpval}}$ with concrete semantics $(\States, \sinit, \flecheRel)$, we refer to the states of~$\States$ as the \emph{concrete states} of~$\valuate{\A}{\twpval{\tpval}{\wpval}}$.
A \emph{run} of~$\valuate{\A}{\twpval{\tpval}{\wpval}}$ is an alternating sequence of concrete states of $\valuate{\A}{\twpval{\tpval}{\wpval}}$ and pairs of edges and delays starting from the initial state $\sinit$ of the form
$\state_0, (\edge_0, d_0), \state_1, \cdots$
with
$i = 0, 1, \dots$, $\edge_i \in \Edges$, $d_i \in \grandrplus$ and $(\state_i , \edge_i , \state_{i+1}) \in \flecheRel$.
\ea{j'ai supprimé les defs sur la reachability mais c'est ptêt une mauvaise idée (?)}

\begin{example}
A concrete execution of the PWTA $\valuate{\A}{\twpval{\tpval}{\wpval}}$ of \cref{example:motivating} with~${\color{magenta} \weight} = 2$\euro, $\wpval( {\color{purple} \wparam} ) = 0.5$\euro, $\tpval({\color{orange}p_{1}})=5$ and $\tpval({\color{pink}p_{2}})=8$ is

	$({\color{blue}l_{1}}, (0,0), (0))\longuefleche{(\styleact{press}, 2)} ({\color{green}l_{2}},(0,0), (2))\longuefleche{(\styleact{press}, 1.5)}({\color{green}l_{2}},(0,1.5), (2.5))\longuefleche{(\styleact{press}, 1)}({\color{green}l_{2}},(0,2.5), (3))\longuefleche{(\styleact{prepare}, 2.5)}({\color{red}l_{3}},(2.5,5), (3))\longuefleche{(\styleact{serve}, 3)}({\color{blue}l_{1}}, (5.5,8), (2.5))$.\ea{j'ai corrigé car tu avais mis ``$,0.5$'' dans le vecteur de coûts, mais on n'a que les coûts constants, pas les coûts paramétrés dans la sémantique ; de même que pour les horloges, on met les valeurs des horloges mais pas celles des paramètres. Pour le reste, super exemple.}

	\LongVersion{Note that no coffee can be served if $\tpval({\color{orange}p_{1}})=8$ and $\tpval({\color{pink}p_{2}})=5$.}
\end{example}

\mr{définition de WCTL avec Time ?}

\LongVersion{
\begin{remark}
	Despite the name of \emph{weights} (justified by our context of measuring costs and damages), our parametric weights are in fact sufficiently expressive to encode parametric (rational-valued) data.
\end{remark}
}
\section{Translation of AFTs to PTAs}\label{section:translation}
\subsection{Overview of the translation}\label{section:overview}

We will model an attack-fault tree using a network of PWTAs that will synchronize along actions (using the usual composition semantics)\ea{alors le problème, c'est que nos PWTA, quoique simples, n'ont pas été définis avant. Faut-il définir la composition parallèle ? (Vu les délais, je réponds non.)}.
Each gate and each leaf (\ie{} BAS or BCF) will be modeled as a PWTA.
Leaves PWTA have a duration and a weight, while gates PWTA store the weight value of their children to forward it to their parents.
Therefore, each gate PWTA maintains its own weight, and its value will be added to that of their parents in case of success (thanks to the parametric weight update).

All gates and leaves PWTAs initially synchronize their start action---referred as \emph{activation} in this paper---, and end with either a success or fail synchronization action. After gates synchronize their start action, they synchronize the start action of their children. %

Intuitively, the process is top-bottom-top: the top level gate PWTA activates its children, which themselves activate their children (if any), and so on until the leaves PWTAs at the bottom of the attack-falt tree.
Once a leaf PWTA terminates, it synchronizes either its success or fail action.
In case of success, the leaf PWTA forwards its weight %
value to its parent, where this value is stored.
When its parent gate PWTA terminates, the gate PWTA synchronizes either a success or a fail action.
In case of success, the gate PWTA forwards its weight value to its parent, and so on until the top-level gate PWTA terminates.

If the top-level PWTA terminates in its success location, the attack is successful.
We apply the reachability synthesis algorithm of PTAs on the success location in the top-level PWTA, that is, we synthesize all valuations for which this location is reachable: this gives us the success conditions of an attack.
The set of constraints on time and weight\LongVersion{ (such as cost for the attacker, damages for the organization\ea{attention là encore on ne parle que d'un unique type de coût}\mr{utilisé du coup pour le deuxieme case study})}
that allowed this attack to be successful are output by this analysis.

As a running example, we consider the attack tree in \cref{ATree} taken from \cite{SYRGKDRS17}.

\subsection{Translation of leaves}

\ea{il faut absolument donner l'exemple de traduction des (d'une) feuille(s). C'est là que les paramètres temporels interviennent. J'ai copié-collé ci-dessous un bout de ton exemple ; réordonne si besoin}

\begin{figure}[tb]
\centering
\scalebox{.8}{
\begin{tikzpicture}[shorten >=1pt, node distance=1.5cm and 3cm, on grid, auto]
 \node[location, initial, initial text={$\styleclock{x}:=0$}, inner sep=1pt] (A)   {$\color{blue}l_{1}$};
 \node[location]          (B)   [right=of A]           {$\color{blue}l_{2}$};
 \path(B.north) node[postaction = {draw, gray, thin},yshift=3mm] (G5) {$\styleclock{x}\leq 5$};
 \node[location, success]          (C)   [above right=of B]           {$l_{3}$};
 \node[location, failure]          (D)   [below right=of B]           {$l_{4}$};

\path[->]
					 (A)  edge   node[above] {\styleact{start}}  node[below] {$\styleclock{x}:=0$}  (B)
					 (B)  edge   node[swap,yshift=4mm] {\begin{tabular}{c}\styleact{success} \\ $\styleclock{x}\geq 5$\\$\weight_{\mbox{parent}} := \weight_{\mbox{parent}} + \weight_{\mbox{leaf}}$\end{tabular}}   (C)
					 (B)	edge   [swap]  node[yshift=4mm]  {\begin{tabular}{c}\styleact{fail} \\ $\styleclock{x}\geq 5$\end{tabular}} (D)

					 ;

\end{tikzpicture}
 }
 \caption{PWTA translation of leaf that can reach the success location in exactly 5 units of time, and of weight~$\weight_{\mbox{leaf}}$} %
 \label{fig:leafPWTA}
 \end{figure}
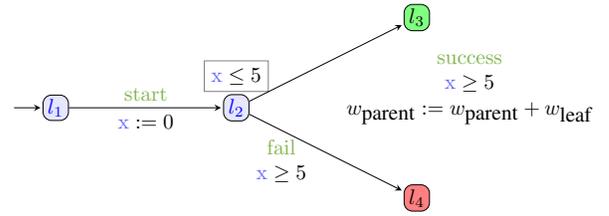

A BAS/BCF is modeled as a PWTA with clocks and weights (see \cref{fig:leafPWTA}). Note that in real life while a BAS needs to success so the attack is possibly successful, a BCF needs to fail in order to propagate a disruption (as in basic component \emph{failure}). However we consider in our models that both BAS and BCF need to reach the success location. %
There is two paths in a BAS/BCF PWTA, one that reaches the success location and one that reaches the fail location. In case of success, its weight is forwarded to and stored in its parent gate.

\LongVersion{
\begin{example}
	The translation of leaf \aft{find\_WLAN} of \cref{ATree} is given in \cref{fig:BAS} in \cref{appendix:fig:BAS}.
\end{example}
}

\subsection{Translation of gates}

Concrete translations of \sandgate{}, \andgate{}, \orgate{} gates are given in \cref{tbl:rules} (yellow locations denote urgency: time cannot elapse).
\LongVersion{%
	We describe them and give examples in the following.
}
Other gates are similar.\ea{mais dans l'idéal (là, pas le temps), on aurait pu les mettre en annexe}

\newcommand{\figsandgatepta}{
  \resizebox{\linewidth}{!}{%
\begin{tikzpicture}[shorten >=1pt, node distance=2cm, on grid, auto]
   \node[location, initial, initial text={$x,y:=0$}, inner sep=1pt] (A)   {$\color{blue}l_{1}$};
   \node[location]          (B)   [right=of A]           {$\color{blue}l_{2}$};
   \node[location]          (C)   [right=of B]           {$\color{blue}l_{3}$};
   \node[location]          (D)   [right=of C]           {$\color{blue}l_{4}$};
	 \node[location]          (E)   [right=of D]           {$\color{blue}l_{5}$};
   \node[location]          (F)   [right=of E]           {$\color{blue}l_{6}$};
   \node[location]          (G)   [right=of F]           {$\color{blue}l_{7}$};
	 \node[location, urgent]          (H)   [right=of G]           {$l_{8}$};
	 \node[location, success]          (I)   [right=of H]           {$l_{s}$};
	 \node[location, urgent]          (J)   [below=of H]           {$l_{9}$};
	 \node[location, failure]          (K)   [below=of I]           {$l_{f}$};
  \path[->]
             (A)  edge   node  [swap]   {\styleact{startSAND}}   (B)
             (B)  edge   node[swap] {\styleact{startA}}   (C)
						 (C)	edge   [swap]  node[above]  {\styleact{successA}} (D)
						 (C)	edge   [swap]  node  {\styleact{failA}} (J)
             (D)  edge [swap]  node {\styleact{startB}} (E)
						 (E)  edge [swap]  node[above] {\styleact{successB}} (F)
						 (E)  edge [swap]  node[yshift=1mm] {\styleact{failB}} (J)
						 (F)  edge [swap]  node {\styleact{startC}} (G)
						 (G)  edge [swap]  node[above] {\styleact{successC}} (H)
						 (G)  edge [swap]  node[xshift=1mm] {\styleact{failC}} (J)
						 (H)  edge [swap]  node[yshift=-3mm] {\begin{tabular}{c}\styleact{successSAND} \\ $\weight_{\mbox{parent}}\pluseq\weight_{\sandgate{}}$\end{tabular}} (I)
						 (J)  edge [swap]  node {\styleact{failSAND}} (K)
						 ;

\end{tikzpicture}
}
}
\newcommand{\figsandgate}{
\definecolor{attackedcolor}{gray}{0.75}
  \resizebox{\linewidth}{!}{%
\begin{tikzpicture}[
node distance=12mm,
rk/.style={rectangle,draw,minimum
height=5mm,fill=black!10, outer sep=0pt,
font=\ttfamily},
zigzag/.style={to path={|- ($(\tikztostart)!.75!(\tikztotarget)$) -| (\tikztotarget)}}]
\node[rk, fill=attackedcolor] (Top) {\sandgate{} gate};
\path(Top.south) node[anchor=east,and1, fill=attackedcolor] (G5) {};
\node[below of=G5, yshift=1mm, rk, fill=attackedcolor] (SISF) {B};
\node[anchor=east, rk,xshift=-5mm, fill=attackedcolor] at
(SISF.west) (PL) {A};
\node[anchor=west ,rk,xshift=5mm, fill=attackedcolor] at
(SISF.east) (SI) {C};

\path[draw] (G5.west) -- ++(0,-4mm) -- (SISF.north -| G5.west);
\path[-]
(PL) edge[zigzag] (G5.input 1)
(SI) edge[zigzag] (G5.input 2)
;
\path[->,line width=1.5pt]
(G5.north) edge (G5.south)
;
\end{tikzpicture}%
}
}
\newcommand{\figandgate}{
\definecolor{attackedcolor}{gray}{0.75}
  \resizebox{\linewidth}{!}{%
\begin{tikzpicture}[
node distance=12mm,
rk/.style={rectangle,draw,minimum
height=5mm,fill=black!10, outer sep=0pt,
font=\ttfamily},
zigzag/.style={to path={|- ($(\tikztostart)!.75!(\tikztotarget)$) -| (\tikztotarget)}}]
\node[rk, fill=attackedcolor] (Top) {\andgate{} gate};
\path(Top.south) node[anchor=east,and1, fill=attackedcolor] (G5) {};
\node[anchor=east, rk,xshift=-5mm, fill=attackedcolor] at
(SISF.west) (PL) {A};
\node[anchor=west ,rk,xshift=5mm, fill=attackedcolor] at
(SISF.east) (SI) {B};

\path[-]
(PL) edge[zigzag] (G5.input 1)
(SI) edge[zigzag] (G5.input 2)
;
\end{tikzpicture}%
}
}
\newcommand{\figandgatepta}{
  \resizebox{\linewidth}{!}{%
\begin{tikzpicture}[shorten >=1pt, node distance=2cm, on grid, auto]
   \node[location, initial, initial text={$x,y:=0$}, inner sep=1pt] (A)   {$\color{blue}l_{1}$};
   \node[location]          (B)   [right=of A]           {$\color{blue}l_{2}$};
   \node[location]          (C)   [right=of B]           {$\color{blue}l_{3}$};
   \node[location]          (D)   [right=of C]           {$\color{blue}l_{4}$};
	 \node[location, urgent]          (J)   [right=of D]           {$l_{7}$};
	 \node[location]          (E)   [above=of J]           {$\color{blue}l_{5}$};
	 \node[location, failure]          (K)   [right=of J]           {$l_{f}$};
	 \node[location]          (H)   [below=of J]           {$\color{blue}l_{8}$};
 \node[location, urgent]          (F)   [right=of K]           {$l_{6}$};
	 \node[location, success]          (G)   [right=of F]           {$l_{s}$};
  \path[->]
             (A)  edge   node  [swap]   {\styleact{startAND}}   (B)
             (B)  edge   node[swap] {\styleact{startA}}   (C)
						 (C)	edge   [swap]  node  {\styleact{startB}} (D)
             (D)  edge [swap]  node[above, xshift=-5mm] {\styleact{successA}} (E)
						 (E)  edge [swap]  node[above, xshift=6mm] {\styleact{successB}} (F)
						 (F)	edge   [swap]  node[yshift=-3mm]  {\begin{tabular}{c}\styleact{successAND} \\ $\weight_{\mbox{parent}}\pluseq\weight_{\andgate{}}$\end{tabular}} (G)
						 (D)  edge [swap]  node {\styleact{successB}} (H)
						 (H)  edge [swap]  node {\styleact{successA}} (F)
						 (D)  edge [swap]  node[yshift=4mm] {\begin{tabular}{c}\styleact{failA} \\ \styleact{failB}\end{tabular}} (J)
						 (E)  edge [swap]  node[above, xshift=4mm] {\styleact{failB}} (J)
						 (H)  edge [swap]  node {\styleact{failA}} (J)
						 (J)  edge [swap]  node {\styleact{failAND}} (K)

						 ;

\end{tikzpicture}
}
}
\newcommand{\figorgate}{
\definecolor{attackedcolor}{gray}{0.75}
  \resizebox{\linewidth}{!}{%
\begin{tikzpicture}[
node distance=12mm,
rk/.style={rectangle,draw,minimum
height=5mm,fill=black!10, outer sep=0pt,
font=\ttfamily},
zigzag/.style={to path={|- ($(\tikztostart)!.75!(\tikztotarget)$) -| (\tikztotarget)}}]
\node[rk, fill=attackedcolor] (Top) {\orgate{} gate};
\path(Top.south) node[anchor=east,or1, fill=attackedcolor] (G5) {};
\node[anchor=east, rk,xshift=-5mm, fill=attackedcolor] at
(SISF.west) (PL) {A};
\node[anchor=west ,rk,xshift=5mm, fill=attackedcolor] at
(SISF.east) (SI) {B};

\path[-]
(PL) edge[zigzag] (G5.input 1)
(SI) edge[zigzag] (G5.input 2)
;
\end{tikzpicture}%
}
}
\newcommand{\figorgatepta}{
  \resizebox{\linewidth}{!}{%
\begin{tikzpicture}[shorten >=1pt, node distance=2cm, on grid, auto]
   \node[location, initial, initial text={$x,y:=0$}, inner sep=1pt] (A)   {$\color{blue}l_{1}$};
   \node[location]          (B)   [right=of A]           {$\color{blue}l_{2}$};
   \node[location]          (C)   [right=of B]           {$\color{blue}l_{3}$};
   \node[location]          (D)   [right=of C]           {$\color{blue}l_{4}$};
	 \node[location, urgent]          (J)   [right=of D]           {$l_{7}$};
	 \node[location]          (E)   [above=of J]           {$\color{blue}l_{5}$};
	 \node[location, success]          (K)   [right=of J]           {$l_{s}$};
	 \node[location]          (H)   [below=of J]           {$\color{blue}l_{8}$};
 \node[location, urgent]          (F)   [right=of K]           {$l_{6}$};
	 \node[location, failure]          (G)   [right=of F]           {$l_{f}$};
  \path[->]
             (A)  edge   node  [swap]   {\styleact{startOR}}   (B)
             (B)  edge   node[swap] {\styleact{startA}}   (C)
						 (C)	edge   [swap]  node  {\styleact{startB}} (D)
             (D)  edge [swap]  node[above, xshift=-3mm] {\styleact{failA}} (E)
						 (E)  edge [swap]  node[above,xshift=3mm] {\styleact{failB}} (F)
						 (F)	edge   [swap]  node  {\styleact{failOR}} (G)
						 (D)  edge [swap]  node {\styleact{failB}} (H)
						 (H)  edge [swap]  node {\styleact{failA}} (F)
						 (D)  edge [swap]  node[yshift=4mm] {\begin{tabular}{c}\styleact{successA} \\ \styleact{successB} \end{tabular}} (J)
						 (E)  edge [swap]  node[above,xshift=6mm] {\styleact{successB}} (J)
						 (H)  edge [swap]  node {\styleact{successA}} (J)
						 (J)  edge [swap]  node [above] {\styleact{successOR}} node[yshift=-3mm, xshift=2mm]  {$\weight_{\mbox{parent}}\pluseq\weight_{\orgate{}}$} (K)

						 ;

\end{tikzpicture}
}
}

\newcommand{\dummyfigure}{\tikz \fill [NavyBlue] (0,0) rectangle node [black] {Figure} (2,2);}
\begin{table*}
		\centering
	\begin{tabularx}{\textwidth}{p{3cm}p{14.3cm}}
			 \toprule
				\figandgate & \figandgatepta \\
				\midrule
				\figsandgate  & \figsandgatepta \\
				\midrule
				\figorgate & \figorgatepta \\
				\bottomrule
		\end{tabularx}
		\caption{Translation rules of \andgate{}, \sandgate{} and \orgate{} gates to PWTA}
		\label{tbl:rules}
\end{table*}

\paragraph{\andgate{}}
recall that an \andgate{} gate is disrupted if all of its children are disrupted.
It activates all of its children then waits for their disruptions regardless of the order of the successes.
At any moment if one fails, the \andgate{} gate fails.
If the success action is synchronized, its parent weight~$\weight_{\mbox{parent}}$ is updated: the weight~$\weight_{\andgate{}}$ carried by the \andgate{} gate is added to~$\weight_{\mbox{parent}}$.

\begin{figure*}[h!]
\footnotesize
\begin{subfigure}[c]{0.5\textwidth}
\includegraphics[width=0.8\textwidth]{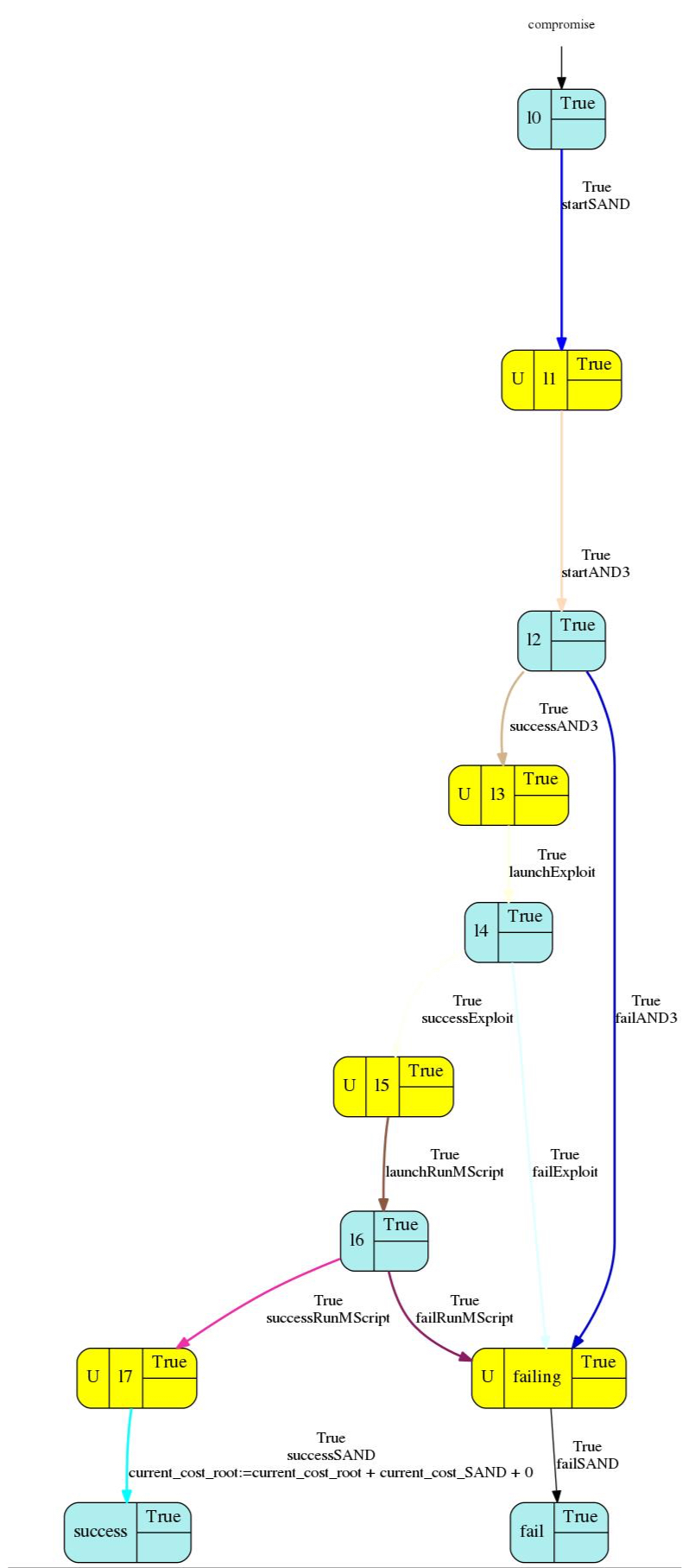}
\caption{Translation of the top-level \sandgate{} gate}
\label{fig:SAND}
\end{subfigure}
\begin{subfigure}[c]{0.5\textwidth}
\includegraphics[width=0.8\textwidth]{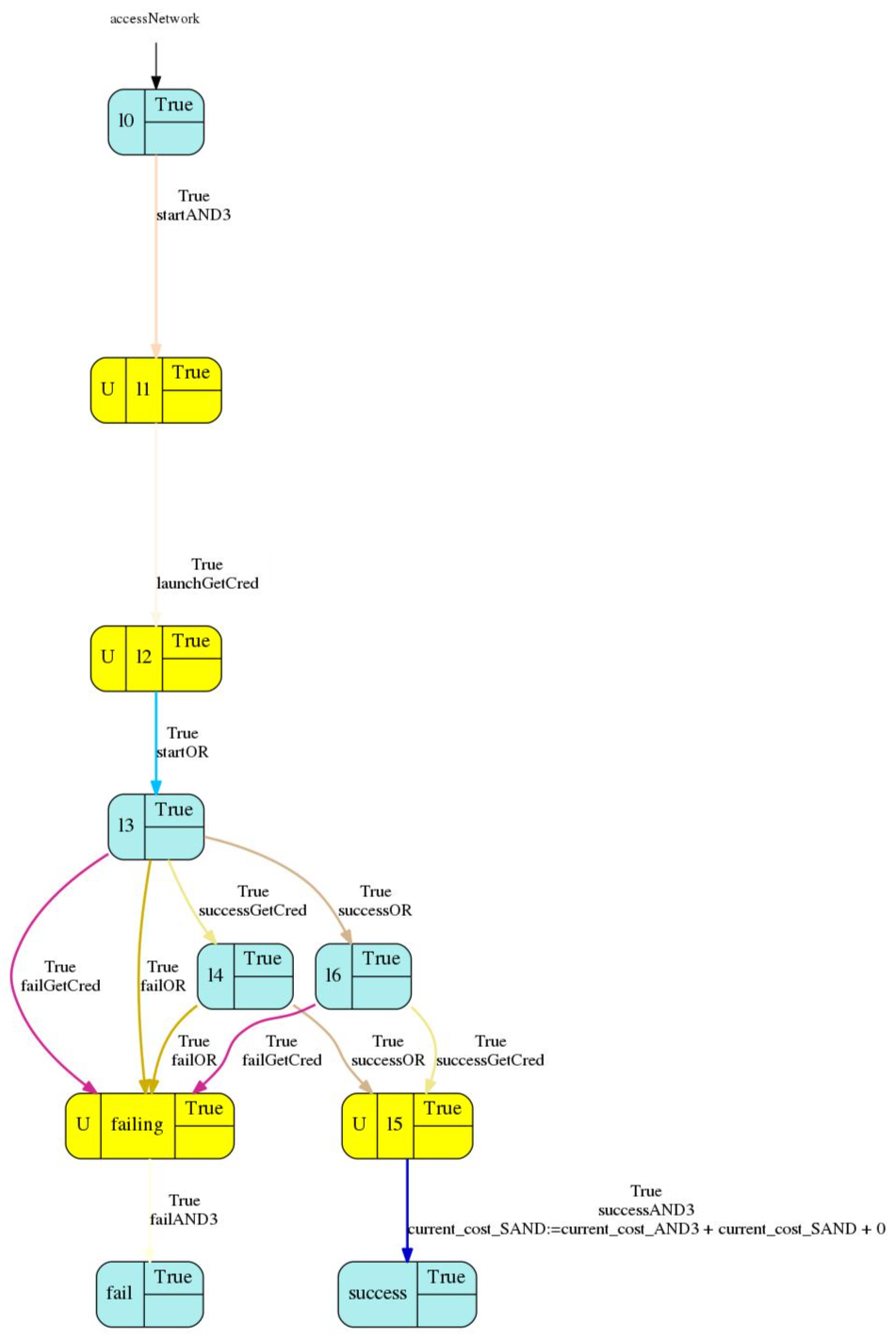}
\caption{Translation of the \andgate{} gate \aft{access\_home\_network}}
\label{fig:AND}
\end{subfigure}
\caption{\sandgate{} and \andgate{} gate}
\end{figure*}

\begin{example}
	We give in \cref{fig:AND} the PWTA corresponding to the \andgate{} gate \aft{access\_home\_network} of \cref{ATree}.
	When all children are activated in the PWTA of \cref{fig:AND}, there are four paths leading to the fail state, while only two (success of the two children in any order) leading to the success state.
	\emph{startAND3} launches the \andgate{} gate \aft{access\_home\_network}. Both children, the BAS \aft{get\_credential} and the \orgate{} gate \aft{gain\_access\_to\_private\_networks} are activated with the synchronization of the actions \emph{launchGetCred} and \emph{startOR}.
	Unlike the \sandgate{} gate, \andgate{} gate waits for any of its child to synchronize a success action.
	If \emph{successGetCred} is synchronized, it then will wait for \emph{successOR} to go to the location success. If \emph{failGetCred} is synchronized, the automaton will go to the location failing. When waiting for the action \emph{successOR}, if \emph{failOR} is synchronized the automaton will also go to the location failing. The other possibility (\emph{successOR} then \emph{successGetCred}) is similar. When in location failing it synchronizes the action \emph{failAND3}, while if going to the location success it will synchronize the action \emph{successAND3}. If the success state is reached, the weight of its parent gate is increased by its own weight.
\end{example}
\paragraph{\sandgate{}}
recall that a \sandgate{} gate is disrupted if all its children from left to right are disrupted sequentially from left to right.
It activates its leftmost child then waits for its success or failure, then activates its second leftmost child and so on. If the rightmost child succeeds, the \sandgate{} gate is disrupted. If one child fails, the \sandgate{} gate fails.
For a \sandgate{} gate modeled as a PWTA with~$n$ children, there is only one path leading to the success state, while there are $n$~paths leading to the fail state (one from each child).
If the success action is synchronized, its parent weight~$\weight_{\mbox{parent}}$ is updated: the weight~$\weight_{\sandgate{}}$ carried by the \sandgate{} gate is added to~$\weight_{\mbox{parent}}$.

\begin{example}
	The top event of the attack tree in \cref{ATree} is a \sandgate{} gate.
	We give the PWTA corresponding to this \sandgate{} in \cref{fig:SAND}.
	It synchronizes the action \emph{startSAND}.
	Then it activates its leftmost child \aft{access\_home\_network} with the action \emph{startAND3}, which is an \andgate{} gate. If the action \emph{successAND3} is synchronized, its second leftmost child is activated with the action \emph{launchExploit}.
	If the action \emph{successExploit} is synchronized, its third and last child is activated with the action \emph{launchRunMScript}. If the action \emph{successRunMScript} is synchronized, the action \emph{successSAND} is synchronized. At any moment, if one of its children fail and an action \emph{failAND3}, \emph{failExploit} or \emph{failRunMscript} is synchronized the automaton goes to the location failing where the action \emph{failSAND} is synchronized. If the success state is reached, the weight of its parent gate is increased by its own weight.
\end{example}
\paragraph{\orgate{}}
\orgate{} gate initially activates all of its children. \orgate{} gate is disrupted if at least one of its children is disrupted, and fails if all of its children fail.
Therefore in the case of two children, one child can fail and the \orgate{} gate still propagates a disruption if the other one succeeds right after. However, if one child succeeds no need to wait for the second one and the success action of the \orgate{} gate is synchronized.
If the success action is synchronized, its parent weight~$\weight_{\mbox{parent}}$ is updated: the weight~$\weight_{\orgate{}}$ carried by the \orgate{} gate is added to~$\weight_{\mbox{parent}}$.
\todo{ici, une explication rapide de la traduction \cref{tbl:rules} (avec les coûts !)}

\LongVersion{
\begin{example}
	The PWTA translating the only \orgate{} of \cref{ATree} is given in \cref{fig:OR} in \cref{appendix:OR}.
\end{example}
}
\subsection{Top-level automaton}

Finally, we need to create an automaton that will activate the first top-event gate of the AFT. We call it rootTA.
This PWTA is the one that starts the chain reaction by activating the top-event PWTA gate, which at its turn will activate its own children and so on. It waits for the success or fail action of this PWTA gate. In case of success, its weight has been updated with the total weight value of the execution forwarded by the top-event gate PWTA.
This bottom-to-top addition stores in the weight \emph{current\_cost\_root} the total weight of the attack.
The rootTA also stores the total time spent since the first activation of the top-event PWTA (using an extra clock and parameter).

\begin{figure}[h!]

\scalebox{.8}{
\begin{tikzpicture}[shorten >=1pt, node distance=1.5cm and 3cm, on grid, auto]
 \node[location, initial, initial text={$\styleclock{x}:=0$}, inner sep=1pt] (A)   {$\color{blue}l_{1}$};
 \node[location]          (B)   [right=of A]           {$\color{blue}l_{2}$};
 \node[location, urgent]          (C)   [above right=of B]           {$l_{3}$};
 \node[location, urgent]          (D)   [below right=of B]           {$l_{5}$};
 \node[location, success]          (E)   [right=of C]           {$l_{4}$};
 \node[location, failure]          (F)   [right=of D]           {$l_{6}$};
\path[->]
					 (A)  edge   node[above] {\styleact{startSAND}}   (B)
					 (B)  edge   node[] {\styleact{successSAND}} (C)
					 (C)  edge   node[below, align=center] {\styleact{successRoot} \\ $ total\_cost = current\_cost\_root$
           \\ $\styleclock{abs\_time} = \styleparam{total\_time}$
					 } (E)
					 	(B)	edge   [swap]  node[]  {\begin{tabular}{c}\styleact{failSAND} \end{tabular}} (D)
					 (D)	edge   [swap]  node[]  {\begin{tabular}{c}\styleact{failROOT} \end{tabular}} (F)

					 ;

\end{tikzpicture}
}

	\caption{The rootTA}
	\label{fig:rootTA}
\end{figure}
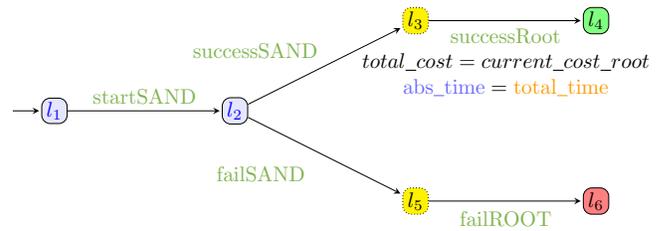
\begin{example}
	We give in \cref{fig:rootTA} the top-level PWTA for the AFT in \cref{ATree}.
	\ea{alors là, il faut refaire la figure en Tikz, l'automate est tout simple ! comme ça, on n'aura pas tous ces stops horribles qui nécessitent des explications alambiquées
	Et détaille en 2-3 lignes max ladite figure}\ea{bon, je l'ai refaite… mais je comprends pas, je ne vois pas de \styleact{failSAND} :/ Pourquoi ?}\mr{en effet; j'ai refait dans imitator et les contraintes n'ont pas changé}
	It is very similar to a leaf PWTA.
	It activates the top-level gate PWTA, then waits for its success or fail action. If the success action is synchronized, its weight has been updated to the total weight value of the execution and is checked against an additional parameter \emph{total\_cost} so \imitator{} outputs this \emph{current\_cost\_root} value. Likewise, the clock~$\styleclock{abs\_time}$ which is never reset since the activation of rootTA is checked against a timing parameter~$\styleparam{total\_time}$. Therefore \imitator{} outputs the total time of execution.
\end{example}
\section{Implementation of the translation}\label{section:implementation}
\subsection{\imitator{}}

\imitator{}~\cite{AFKS12} is a parametric model checker taking as input an extension of networks parametric timed automata extended with synchronization, stopwatches and discrete variables.
\imitator{} supports global (shared) discrete rational-valued variables, that can be either \emph{concrete} (in which case they are syntactic sugar for an unbounded number of locations), or \emph{symbolic}, in which case they can be updated to or compared with parameters.
While \imitator{} technically considers a single type of parameters (where symbolic variables can be compared or even updated to \emph{timing} parameters), our weight parameters are never compared to timing parameters, and this setting can be considered as a subclass of the \imitator{} expressiveness.

\imitator{} implements several synthesis algorithms, notably reachability synthesis (\EFsynth{}), that attempts to synthesize all parameter valuations for which a given location is reachable---which is the algorithm we use here.
\imitator{} relies on the symbolic semantics of parametric timed automata (see \eg{} \cite{ACEF09,JLR15}), where symbolic states are made of a discrete location, and a constraint over the clocks and parameters.
The weight parameters are added to this symbolic semantics in a straightforward manner, with symbolic states enriched with linear constraints over weight parameters.

Note that, while parametric timed automata are highly undecidable (see \cite{Andre19STTT} for a survey), and while our parametric extension adds a new layer of complexity, all analyses terminate with an exact result (sound and complete) because our models are \emph{acyclic}: our AFTs are trees, and their translation yields structurally acyclic PWTAs.
As a consequence, the symbolic semantics of these PWTAs can be represented as a finite structure, and the analysis is guaranteed to terminate.

\subsection{Translation from AFTs to PWTAs}

The translation from AFTs to PWTAs was implemented within the framework of ATTop \cite{SYRGKDRS17}.
The existing software ATTop can take as input a \galileo{} formatted file.
\LongVersion{This format is pretty easy to use and to understand.
}The code in \cref{code:example-galileo} expresses an attack-fault tree of one \orgate{} gate named \aft{A}, with two children \aft{B} and \aft{C}. The BAS \aft{B} takes between~50 and~100 units of time to terminate, and costs~\$50 to the attacker. The BAS \aft{C} takes between~30 and~70 units of time to terminate, and costs~\$30 to the attacker.

\begin{figure}[tbh]
\begin{lstlisting}[style=galileo]
toplevel "A";
"A" or "B" "C";
"B" mintime=50 maxtime=100 cost=50;
"C" mintime=30 maxtime=70 cost=30;
\end{lstlisting}
\caption{Example of \galileo{} attack-tree}
\label{code:example-galileo}
\end{figure}

ATTop takes as an input a \galileo{} file and parses it to represent it as an attack-fault tree meta-model (ATMM) (see \cite[Section 3]{SYRGKDRS17}\LongVersion{, and \cref{fig:tool} in \cref{appendix:tool}} for a screenshot of the tool).
\LongVersion{%
	Then, different translations are available: one quite interesting is the translation into an \uppaal{} file, for instance a network of stochastic timed automata \cite{KS17}. ATTop takes the ATMM and translates it in its \uppaal{} meta-model, then serializes it into an \uppaal{} formatted file.
}
In our approach we directly translate the representation of the ATMM into an \imitator{} formatted file, using the Epsilon Generation Language (EGL) \cite{RPKP08}.
This translation is a very efficient way to obtain AFTs modeled using PWTAs: designing manually a PWTA model from an AFT is very tedious to achieve, while defining an AFT within the \galileo{} syntax is simple.

Once the PWTA obtained, we synthesize using \imitator{} all parameter valuations for which the success location of the rootTA can be reached (using \EFsynth{}).
These sets of parameter values will help us to determine attack and fault scenarios in the following section.

\section{Case studies}\label{section:casestudies}

As a proof of concept, we apply our approach to an attack tree from the literature and an original attack-fault tree.\ea{Mathias, ça te va comme formulation ?}\mr{d'accord original pour nouveau c'est ça? je n'avais pas compris avant mettre l'AFT original}
Experiments were conducted with \imitator{} 2.10.4 ``Butter Jellyfish'',\footnote{%
	Sources, binaries, models and results are available at \url{https://www.imitator.fr/static/ACSD19PAT/}
}
on a 2.4\,GHz Intel Core i5 processor with 2 GiB of RAM in a VirtualBox environment.
Computation times of parameter values ranges from 1 to~9 seconds with four parameters.

\subsection{Compromising an IoT device}

We apply our approach to the AFT depicted in \cref{ATree}\LongVersion{ taken from \cite{SYRGKDRS17}}.
We choose to parametrize the cost of finding a LAN access point ($\styleparam{CostFindLAN\_AP}$) and the maximum amount of time to break WPA keys ($\styleparam{tMax\_Break}$) of the AFT\dl{AFT?}\mr{en l'occurence c'est un AT mais mieux vaut mettre AFT pour la cohérence du papier ?}.
This configuration will describe which attack (WLAN or LAN) is smarter for the attacker, depending on their resources: finding a LAN access point can be difficult depending on the infrastructure security and perhaps social engineering is needed. However, if the attacker does not have enough resources but a large amount of time (s)he can spend time trying to break WPA keys.
\imitator{} computes several constraints on these parameters such that the attack is successful.

Different constraints are possible representing possible time and weight values \st{} an attack is possible. This is represented as a disjunction of conjunctions of constraints on parameters. For instance it can be a quick but very costly attack, or a long but cheap one; therefore different attack and fault scenarios appear.\ea{tu veux dire que \imitator{} retourne une disjonction de contraintes convexes, et que chacune de ces contraintes peut être vue comme une condition pour réussir l'attaque, et donc peut consistuer un attacker profile ?}
The conjunction of constraints
\begin{gather*}
	2*\styleparam{tMax\_Break} \geq 23
	\land \styleparam{CostFindLAN\_AP} \geq 0
	\\
	\land \styleparam{CostFindLAN\_AP} + 180 = \styleparam{total\_cost}
	\land 2*\styleparam{total\_time} = 23
	\\
\end{gather*}
 represents an attack that is very expensive for the attacker: indeed, the total cost of the attack is at least \$$180$ and fully depends on the cost of finding a LAN access point. However, the time spent on the attack is negligible and fixed ($11.5$h).

In opposition, the constraint
\begin{gather*}
	2*\styleparam{tMax\_Break} + 3 \geq 2*\styleparam{total\_time}
	\\
	\land \styleparam{CostFindLAN\_AP} \geq 0
	\\
	\land 2*\styleparam{total\_time} \geq 23
	\land \styleparam{total\_cost} = 232
	\\
\end{gather*}
  shows a an attack that will last at least~$11.5$h---that is, the attacker does not exactly knows when (s)he will break the WPA keys depending for instance of her/his computation power---but with a fixed cost of \$$232$.\dl{il y a quand même une borne supérieure pour total time, certes dépendante d'un autre paramètre}\mr{qui lui est non borné non?}\dl{ah ok, vu comme ça !}.

	Contrarily to our initial intuition, the cost of this second attack can be high above the first one, as breaking the WPA keys is quite costly (\$$80$) in opposition with finding a LAN access point. A smart attacker could choose, regardless of their time and resources the first attack through LAN access point.

\begin{figure}[tbh]
  \begin{center}
	\hspace{-3em}
    \includegraphics[width=0.53\textwidth]{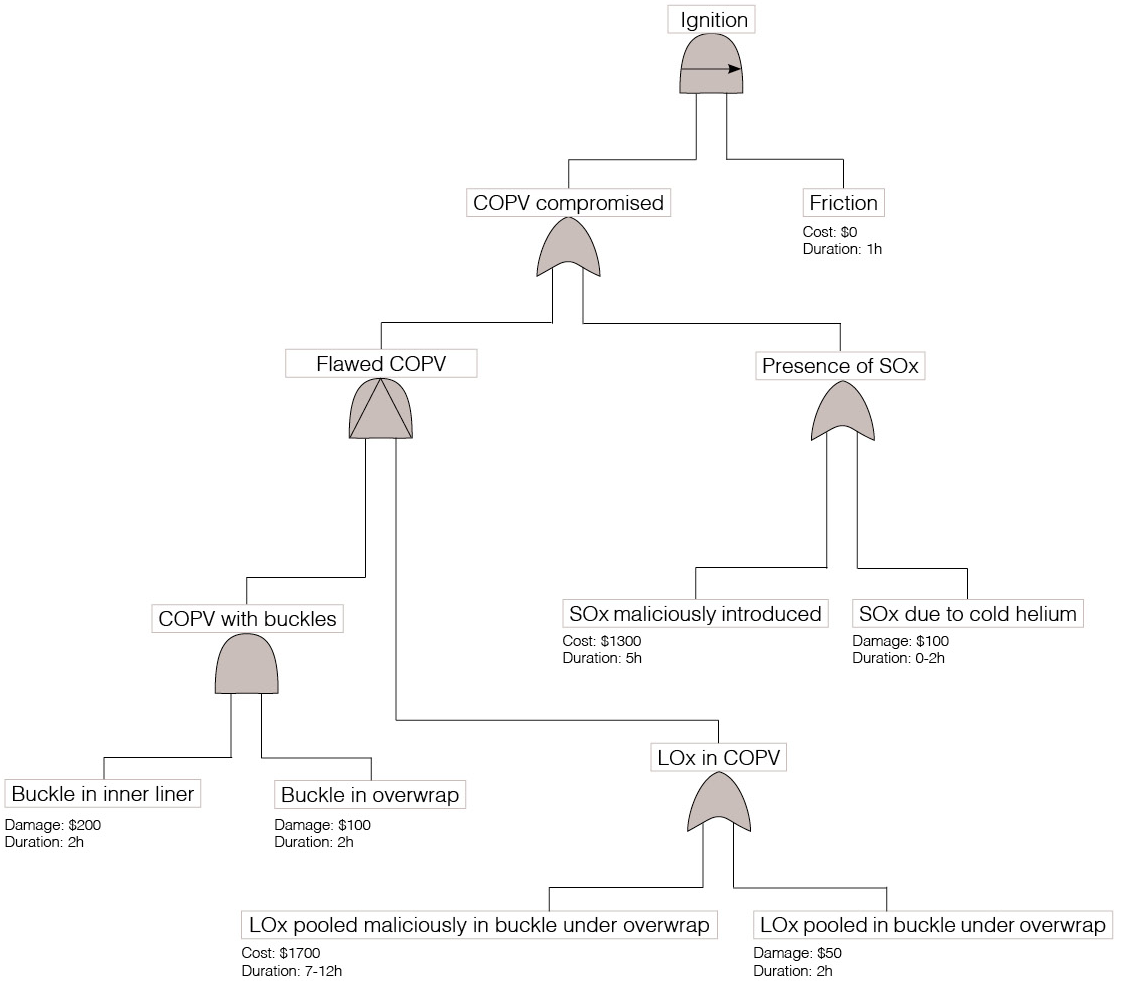}
    \caption{AFT of SpaceX rocket explosion}
    \label{fig:spacex}
  \end{center}
\end{figure}
\subsection{SpaceX rocket Falcon 9 explosion}

Our second case study is an adaptation of the anomaly investigation that followed the explosion of SpaceX rocket Falcon 9 in september 2016\footnote{%
	SpaceX anomaly update, \url{https://www.spacex.com/news/2016/09/01/anomaly-updates}
}.
The AFT in \cref{fig:spacex} depicts the different configurations that can eventually end up with the explosion.
The objective of this case-study is to show that the explosion is more likely to be accidental, due to the expensiveness of the BAS for the attacker who could attempt a sabotage.

The rocket carries a helium tank with three composite overwrapped pressure vessels (COPVs) inside. One COPV possibly had a manufacturing defect and buckles in its liner and the carbon overwrap (\andgate{} gate). Afterwards (\pandgate{} gate) liquid oxygen (LOx) can pool in these buckles and become trapped when pressurized under the carbon overwrap, resulting in a flawed COPV.
An other possibility is the presence of solid oxygen (SOx) either due to the loading temperature of helium or placed here intentionally by an attacker (\orgate{} gate).

These two configurations result in a compromised COPV. When the COPV is compromised, a friction due to take-off tests can start the rocket ignition (\sandgate{} gate).

BCFs have a duration representing the time taken until the component failure. Damage is the cost for the organization for having built a defective component, or the cost induced when the component has failed.
BASs have a cost for the attacker to perform the attack, and a duration for the attack to be successful.

We choose to parametrize the damages induced to the manufacturing facility by $\styleparam{damage\_BuckleInInnerLiner}$, and the cost of pooling solid oxygen near the COPV, $\styleparam{cost\_SOXmaliciouslyIntroduced}$.

The constraint
\begin{gather*}
	13 \geq \styleparam{total\_time} \geq 8
	\\
	\land \styleparam{cost\_SOXmaliciouslyIntroduced} \geq 0
	\\
	\land \styleparam{total\_damages} \geq 100
	\\
	\land \styleparam{damage\_BuckleInInnerLiner} + 100 = \styleparam{total\_damages}
	\\
	\land \styleparam{total\_cost} = 1700
\end{gather*}
represents the attack using the malicious introduction of LOx between the inner liner and the carbon overwrap of the COPV. Clearly this attack is very costly (\$$1700$) and assumes the presence of these buckles. It is highly prejudicial to SpaceX as the company may want to investigate the manufacturing facility that produces COPV components.

The other attack, represented by the constraint
\begin{gather*}
	\styleparam{total\_time} = 6
	\\
	\land \styleparam{cost\_SOXmaliciouslyIntroduced} \geq 0
	\\
	\land \styleparam{total\_damages} = 0
	\\
	\land \styleparam{damage\_BuckleInInnerLiner} \geq 0
	\\
	\land \styleparam{total\_cost} = \styleparam{cost\_SOXmaliciouslyIntroduced}
\end{gather*}
shows that the cost of the attack is equal to the cost of introducing SOx near the COPV. The higher is the parameter $\styleparam{cost\_SOXmaliciouslyIntroduced}$, the higher is the cost of the attack. We may assume this cost is high enough as SpaceX surely secured its launch complex. Otherwise, an efficient counter-measure would be to find means to increase this cost for the attacker.

The constraint
\begin{gather*}
	 \styleparam{cost\_SOXmaliciouslyIntroduced} \geq 0
	\\
	\land \styleparam{total\_damage} \geq 150
	\\
	\land \styleparam{damage\_BuckleInInnerLiner} +150 \geq \styleparam{total\_damage}
	\\
	\land \styleparam{total\_time} = 3
	\land \styleparam{total\_cost} = 0
\end{gather*}
represents the fact that buckles in the inner liner and in the carbon overwrap of the COPV, and then LOx pooled under the overwrap, lead to a complete failure of the system, \ie{} the rocket explodes. In this scenario, there is in all likelihood no attacker. However, the damages for the manufacturing facility can be huge if it is flawed: SpaceX should probably investigate in their manufacturing facilities in order to prevent the production of other flawed components.

Finally, the constraint
\begin{gather*}
 \styleparam{cost\_SOXmaliciouslyIntroduced} \geq 0
	\\
	\land \styleparam{total\_damage} = 100
	\\
	\land \styleparam{damage\_BuckleInInnerLiner} \geq 0
	\\
	\land 3 \geq \styleparam{total\_time} \geq 1
	\land \styleparam{total\_cost} = 0
\end{gather*}
shows that the explosion can be provoked by the presence of SOx due to cold helium. This case is possible without any attacker or component failure and is therefore fully accidental. No damages are caused to SpaceX (excepted the cost of the unusable rocket) or its suppliers.

These scenarios indicate that the rocket explosion is more likely to be accidental, as the cost in both scenarios where there is an attacker is very high. However, the worst case indicates that SpaceX should investigate their production lines to prevent other flawed components, as well as the presence of an attacker.

\section{Conclusion}\label{section:finalremarks}

We addressed the problem of formalizing attack-fault trees in a more abstract framework allowing to cope with parametric timings, costs and damages.
We defined and implemented a translation from attack-fault trees to PWTAs (a new extension of PTAs) that can be analyzed using the \imitator{} model-checker.
This translation allows us to define easily an AFT using the \galileo{} syntax, and obtain as an output this AFT modeled with PWTAs.
Using \imitator{}, we synthesize all parameter values such that there is a successful attack and/or a system failure.
Finally, obtaining a disjunction of convex sets of parameter values allows us to define different attack and fault scenarios\ea{est-ce qu'on garde toujours ça…?}\dl{moi je trouve ça chouette alors c'est bien de le mettre en avant: peut-être ``obtaining a disjunction of convex sets of parameter values'' est plus précis?}. Therefore it helps selecting the most plausible scenario and the most efficient counter-measures.

\paragraph*{Future works}
In this paper, we only considered three parameters: timing, cost and damage parameters.
However, it is trivial to split these parameters into more precise ones, such as human damages (health and insurance) and material damages caused by the attacker or the failure of the system: an attack can be cheap for the attacker but inflict many kind of damages to the organization, as in our SpaceX case study.
Thanks to the vector of weights defined in our PWTAs, this would be immediate to consider in our framework and implementation.

Moreover, extending our framework to attack-defense trees~\cite{KMRS14,GHLLO16} is also on our agenda.

Finally, adding probabilities in order to create probabilistic parametric attack-fault trees will be an interesting and challenging future work. Indeed, in our SpaceX rocket case study adding probabilities to the manufacturing defects of the COPV on top of the damages inflicted to the company would strengthen considerably our formalism.

\section*{Acknowledgements}
We are thankful to %
Stefano Schivo and %
Enno J.J.\ Ruijters from Twente University for their helpful advices concerning meta-models and ATTop.

\todo{Mathias (non urgent !! donc plutôt à la fin), peux-tu bien reformatter tes références ? Il faudrait qu'elles soient toutes présentées dans le bibtex en suivant mes habitudes (qui, à mon avis sont bonnes :p) ; tu peux t'inspirer de \cite{AFKS12, AHV93} par exemple. En général, je pars de DBLP, et je reformatte à la main le champ \texttt{booktitle} en ajoutant \texttt{longbooktitle}, \texttt{location}, \texttt{confdates}.
Assure toi que tous les prénoms d'auteurs soient là.
Les noms de revues devraient être non abrégés (DBLP a parfois des habitudes un peu énervantes).
Attention aussi aux noms en majuscule dans les titres dans ton bibtex, par exemple ``DAG-based'' qui devrait être ``\{DAG\}-based''; idem pour ``\{OWASP\}. \{CISO\} \{A\}pp\{S\}ec''; and so on.
}
\ea{bon je me suis farci ça (et ai gagné une colonne entière !)… prochaine fois, c'est toi qui t'en charge}

\ifdefined\VersionLong
	\newcommand{\ENTCS}{Electronic Notes in Theoretical Computer Science}
	\newcommand{\FMSD}{Formal Methods in System Design}
	\newcommand{\IJFCS}{International Journal of Foundations of Computer Science}
	\newcommand{\IJSSE}{International Journal of Secure Software Engineering}
	\newcommand{\JLAP}{Journal of Logic and Algebraic Programming}
	\newcommand{\JLC}{Journal of Logic and Computation}
	\newcommand{\LNCS}{Lecture Notes in Computer Science}
	\newcommand{\RESS}{Reliability Engineering \& System Safety}
	\newcommand{\STTT}{International Journal on Software Tools for Technology Transfer}
	\newcommand{\TCS}{Theoretical Computer Science}
	\newcommand{\ToPNoC}{Transactions on Petri Nets and Other Models of Concurrency}
	\newcommand{\TSE}{IEEE Transactions on Software Engineering}
	\renewcommand*{\bibfont}{\small}
	\printbibliography[title={References}]
\else
	\bibliographystyle{IEEEtran} %
	\newcommand{\ENTCS}{ENTCS}
	\newcommand{\FMSD}{FMSD}
	\newcommand{\IJFCS}{IJFCS}
	\newcommand{\IJSSE}{IJSSE}
	\newcommand{\JLAP}{JLAP}
	\newcommand{\JLC}{JLC}
	\newcommand{\LNCS}{LNCS}
	\newcommand{\RESS}{RESS}
	\newcommand{\STTT}{STTT}
	\newcommand{\TCS}{TCS}
	\newcommand{\ToPNoC}{ToPNoC}
	\newcommand{\TSE}{TSE}
	\bibliography{pat}
\fi
\ifdefined\VersionLong
\newpage
\appendix
\subsection{Translation of leaf \aft{find\_WLAN}}\label{appendix:fig:BAS}
\begin{figure}[htb!]
	\includegraphics[width=0.8\columnwidth]{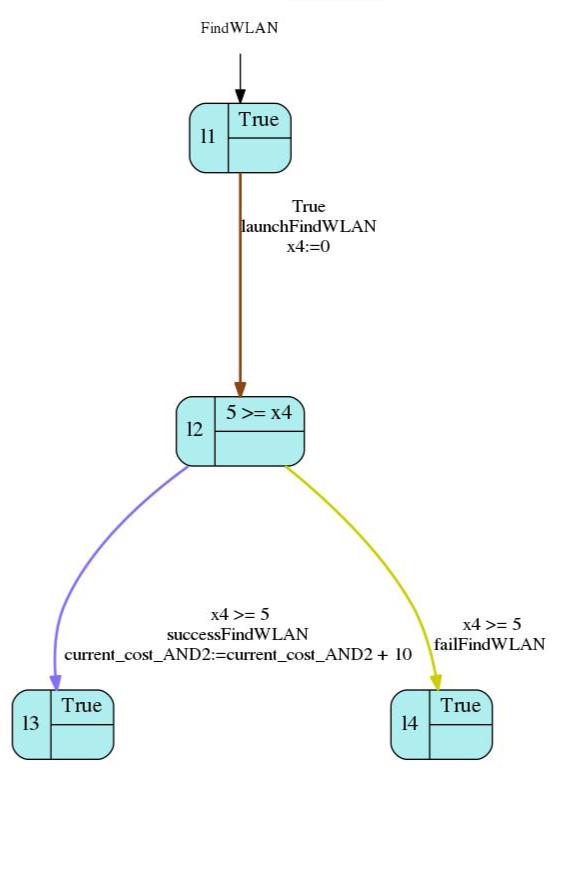}
	\caption{BAS translation of {\ttfamily find\_WLAN}}
	\label{fig:BAS}
\end{figure}

The translation of leaf \aft{find\_WLAN} of \cref{ATree} is given in \cref{fig:BAS} in \cref{appendix:fig:BAS}.\todo{pas prioritaire : mais le refaire en Tikz pour cet automate, ce serait nettement plus clair ! (et prendrait ptêt moins de place ?)}
To express the leaf \aft{find\_WLAN} we use four locations, one clock~$\clock 4$. The first step is to activate the basic attack step using the synchronization action \emph{launchFindWLAN}. Once activated, and at most five units of time after (modeled by the invariant~$5 \geq \clock 4$ and the guard~$\clock 4\geq 5$) it can either success with the action \emph{successFindWLAN} or fail with the action \emph{failFindWLAN}. If the success state is reached, the weight of its parent gate is increased by its own weight~$10$.

\subsection{Example of \orgate{} translation}\label{appendix:OR}

	The PWTA translating the only \orgate{} of \cref{ATree} (given in \cref{fig:OR}) activates all of its children, then waits for one to succeed, regardless of the order.
	Afterwards, whatever happens leads to the success state.
	If one child fails, then the other has to succeed, otherwise the \orgate{} fails.
	Therefore there is six possible paths to the success state, while there is two paths to the fail state (failure of both children in any order).
	\emph{startOR} launches the \orgate{} gate \aft{gain\_access\_to\_private\_networks} which activates its two children using the actions \emph{startAND1} and \emph{startAND2} which activates the \andgate{} \aft{access\_LAN} and \andgate{} \aft{access\_WLAN}. Only one action \emph{successAND1} or \emph{successAN2} is needed to be synchronized so the automaton goes to the location success regardless of which action is synchronized afterwards. Then it synchronizes the action \emph{successOR}. If at first the action \emph{failAND1} (resp.\ \emph{failAND2}) is synchronized, then \emph{successAND2} (resp.\ \emph{successAND1}) has to be synchronized afterwards in order to reach the location success. Otherwise, if \emph{failAND2} (resp.\ \emph{failAND1}) is synchronized, the automaton will go to the location failing and then synchronize the action \emph{failOR}. If the success state is reached, the weight of its parent gate is increased by its own weight.

\begin{figure}[h!]
\footnotesize
\includegraphics[width=0.8\columnwidth]{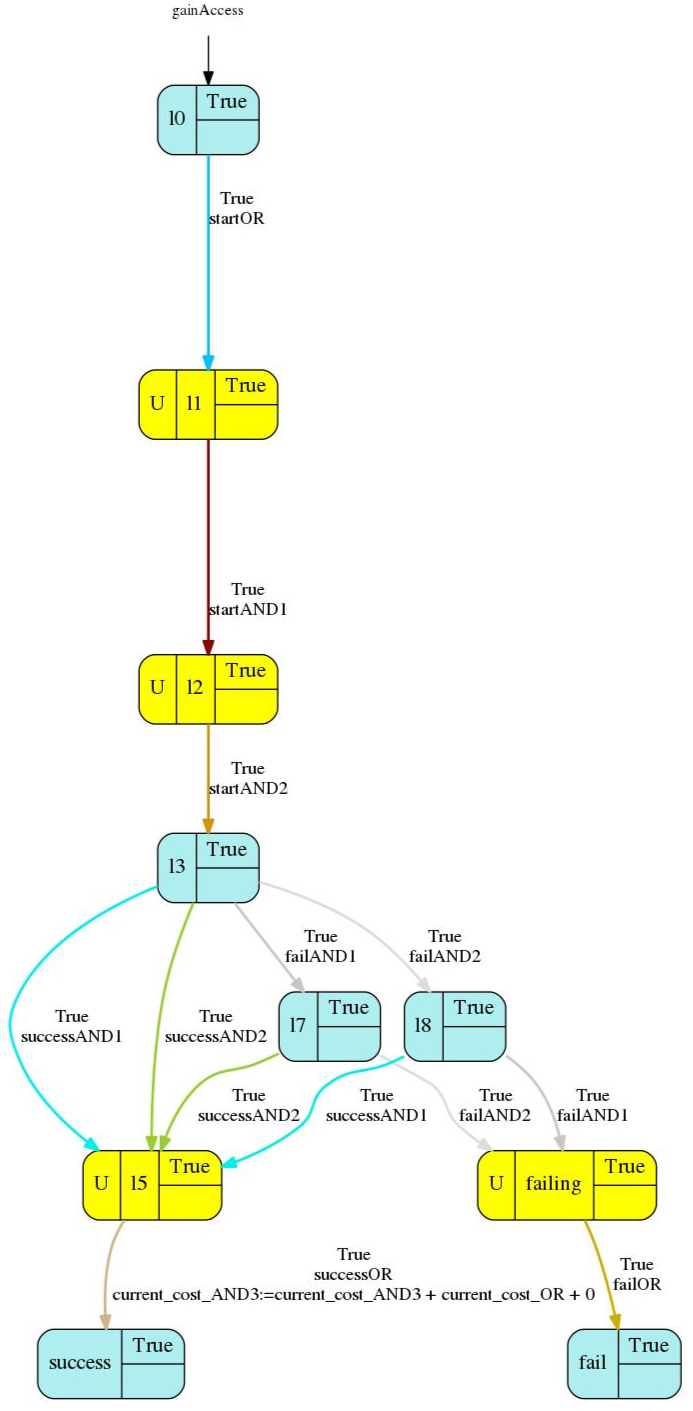}
\caption{Translation of the \orgate{} gate \aft{gain\_access\_to\_private\_networks}}
\label{fig:OR}
\caption{\orgate{}}
\end{figure}
\subsection{Screenshot of the tool ATTop}\label{appendix:tool}
\begin{figure*}[h!]
\footnotesize
\includegraphics[width=\textwidth]{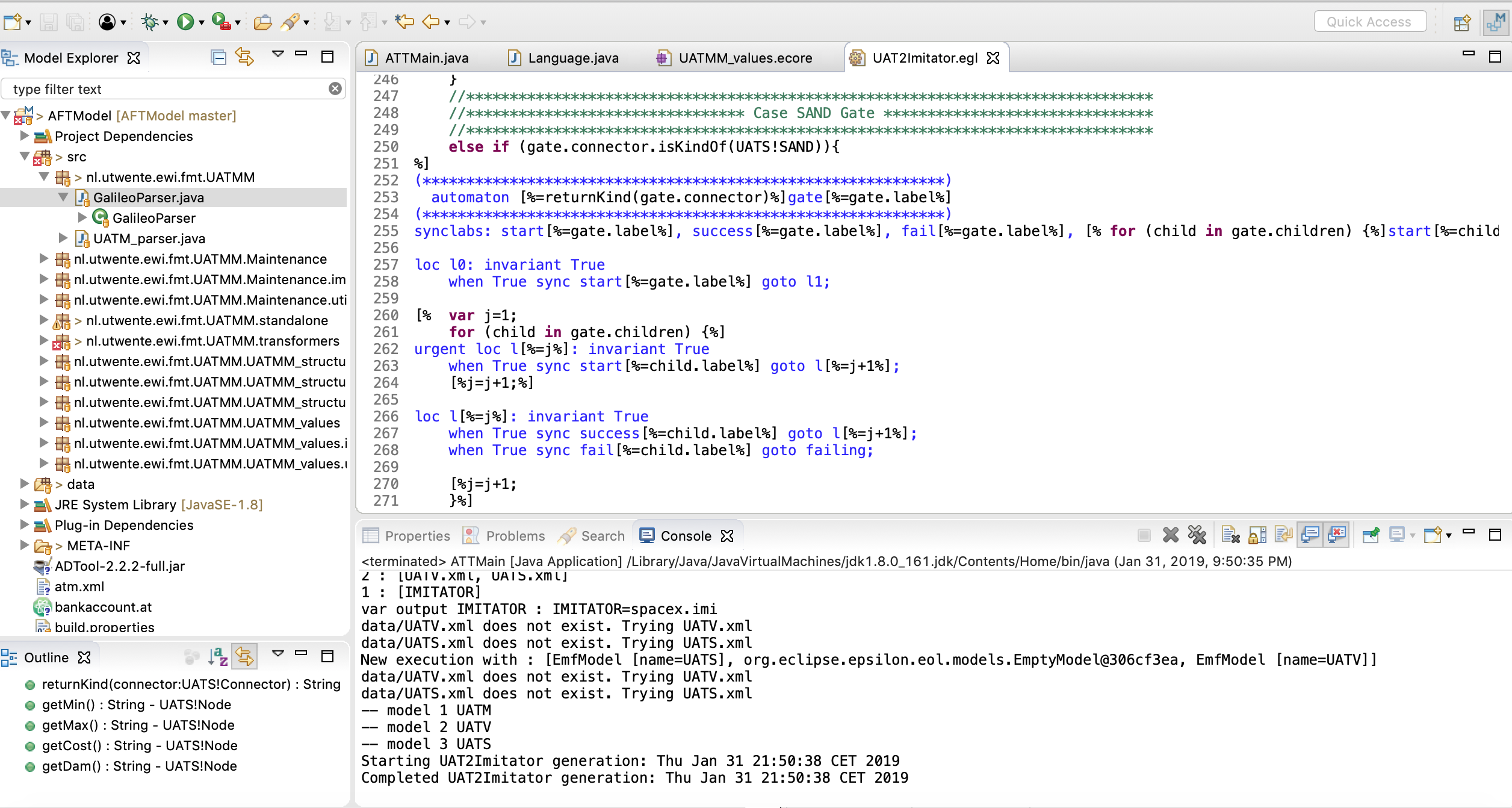}
\caption{Screenshot of the tool ATTop after the translation of the SpaceX AFT}
\label{fig:tool}
\end{figure*}

\fi

\end{document}

%% file: dft-tikz.tex
\usetikzlibrary{calc,arrows,shapes.gates.logic.US,trees,positioning,automata,circuits,fit,shapes.geometric,shapes,intersections,shapes.symbols,shadows,decorations.pathmorphing,decorations,decorations.markings,matrix}

\makeatletter

\pgfdeclareshape{spare}{
  \savedanchor\northeast{%
    \pgfmathsetlength\pgf@x{\pgfshapeminwidth}%
    \pgfmathsetlength\pgf@y{\pgfshapeminheight}%
    \pgf@x=0.5\pgf@x
    \pgf@y=0.5\pgf@y
  }
  \savedanchor\southwest{%
    \pgfmathsetlength\pgf@x{\pgfshapeminwidth}%
    \pgfmathsetlength\pgf@y{\pgfshapeminheight}%
    \pgf@x=-0.5\pgf@x
    \pgf@y=-0.5\pgf@y
  }
  \inheritanchorborder[from=rectangle]

  \anchor{center}{\pgfpointorigin}
  \anchor{north}{\northeast \pgf@x=0pt}
  \anchor{east}{\northeast \pgf@y=0pt}
  \anchor{south}{\southwest \pgf@x=0pt}
  \anchor{west}{\southwest \pgf@y=0pt}
  \anchor{north east}{\northeast}
  \anchor{north west}{\northeast \pgf@x=-\pgf@x}
  \anchor{south west}{\southwest}
  \anchor{south east}{\southwest \pgf@x=-\pgf@x}
  \anchor{text}{
    \pgfpointorigin
    \advance\pgf@x by -.5\wd\pgfnodeparttextbox%
    \advance\pgf@y by -.5\ht\pgfnodeparttextbox%
    \advance\pgf@y by +.5\dp\pgfnodeparttextbox%
  }

  \anchor{O}{ %
    \pgf@process{\northeast}%
    \pgf@x=0pt%
  }
  \anchor{P}{ %
    \pgf@process{\northeast}%
    \pgf@x=-.5\pgf@x%
    \pgf@y=-\pgf@y%
  }
  \anchor{SA}{ %
    \pgf@process{\northeast}%
    \pgf@x=.1\pgf@x%
    \pgf@y=-\pgf@y%
  }
  \anchor{SB}{ %
    \pgf@process{\northeast}%
    \pgf@x=.3\pgf@x%
    \pgf@y=-\pgf@y%
  }
  \anchor{SC}{ %
    \pgf@process{\northeast}%
    \pgf@x=.5\pgf@x%
    \pgf@y=-\pgf@y%
  }
  \anchor{SD}{ %
    \pgf@process{\northeast}%
    \pgf@x=.7\pgf@x%
    \pgf@y=-\pgf@y%
  }
  \anchor{SE}{ %
    \pgf@process{\northeast}%
    \pgf@x=.9\pgf@x%
    \pgf@y=-\pgf@y%
  }
  \anchor{BOX}{ %
    \pgf@process{\northeast}%
    \pgf@x=\pgf@x%
    \pgf@y=-\pgf@y%
  }
  \anchor{LINE}{
    \pgf@process{\northeast}%
    \pgf@x=-1\pgf@x%
    \pgf@y=.\pgf@y%
  }
  \backgroundpath{
    \pgfpathrectanglecorners{\southwest}{\northeast}
    \pgf@anchor@spare@LINE
    \pgf@xa=\pgf@x \pgf@ya=\pgf@y
    \pgf@xb=\pgf@x \pgf@yb=\pgf@y
    \advance\pgf@xb by -2\pgf@x
    \pgfpathmoveto{\pgfpoint{\pgf@xa}{\pgf@ya}}
    \pgfpathlineto{\pgfpoint{\pgf@xb}{\pgf@ya}}
    \pgf@anchor@spare@BOX
    \pgf@xa=\pgf@x \pgf@ya=\pgf@y
    \pgf@xb=\pgf@x \pgf@yb=\pgf@y
    \pgf@xc=\pgf@x \pgf@yc=\pgf@y
    \advance\pgf@ya by -.\pgf@y
    \advance\pgf@yb by -.8\pgf@y
    \advance\pgf@xb by -1\pgf@x
    \pgfpathmoveto{\pgfpoint{\pgf@xa}{\pgf@ya}}
    \pgfpathlineto{\pgfpoint{\pgf@xa}{\pgf@yb}}
    \pgfpathlineto{\pgfpoint{\pgf@xb}{\pgf@yb}}
    \pgfpathlineto{\pgfpoint{\pgf@xb}{\pgf@ya}}
    \pgfclosepath

    \begingroup
    \tikzset{spare/port labels} %
    \tikz@textfont

    \pgf@anchor@spare@O
    \pgftext[top,at={\pgfpoint{\pgf@x}{\pgf@y}},y=-\pgfshapeinnerysep]{}

    \pgf@anchor@spare@P 
    \pgftext[bottom,at={\pgfpoint{\pgf@x}{\pgf@y}},y=\pgfshapeinnerysep]{}
    
    \pgf@anchor@spare@SA
    \pgftext[bottom,at={\pgfpoint{\pgf@x}{\pgf@y}},y=\pgfshapeinnerysep]{}
    
    \pgf@anchor@spare@SB
    \pgftext[bottom,at={\pgfpoint{\pgf@x}{\pgf@y}},y=\pgfshapeinnerysep]{}
    
    \pgf@anchor@spare@SC
    \pgftext[bottom,at={\pgfpoint{\pgf@x}{\pgf@y}},y=\pgfshapeinnerysep]{}
    
    \pgf@anchor@spare@SD
    \pgftext[bottom,at={\pgfpoint{\pgf@x}{\pgf@y}},y=\pgfshapeinnerysep]{}
    
    \pgf@anchor@spare@SE
    \pgftext[bottom,at={\pgfpoint{\pgf@x}{\pgf@y}},y=\pgfshapeinnerysep]{}
    \endgroup
  }
}

\tikzset{add font/.code={\expandafter\def\expandafter\tikz@textfont\expandafter{\tikz@textfont#1}}} 

\tikzset{spare/port labels/.style={font=\sffamily\scriptsize}}
\tikzset{every spare node/.style={draw,minimum height=1cm,minimum 
width=1.5cm,inner sep=1mm,outer sep=0pt,cap=round,add 
font=\sffamily}}

\pgfdeclareshape{spareM}{
  \savedanchor\northeast{%
    \pgfmathsetlength\pgf@x{\pgfshapeminwidth}%
    \pgfmathsetlength\pgf@y{\pgfshapeminheight}%
    \pgf@x=0.5\pgf@x
    \pgf@y=0.5\pgf@y
  }
  \savedanchor\southwest{%
    \pgfmathsetlength\pgf@x{\pgfshapeminwidth}%
    \pgfmathsetlength\pgf@y{\pgfshapeminheight}%
    \pgf@x=-0.5\pgf@x
    \pgf@y=-0.5\pgf@y
  }
  \inheritanchorborder[from=rectangle]

  \anchor{center}{\pgfpointorigin}
  \anchor{north}{\northeast \pgf@x=0pt}
  \anchor{east}{\northeast \pgf@y=0pt}
  \anchor{south}{\southwest \pgf@x=0pt}
  \anchor{west}{\southwest \pgf@y=0pt}
  \anchor{north east}{\northeast}
  \anchor{north west}{\northeast \pgf@x=-\pgf@x}
  \anchor{south west}{\southwest}
  \anchor{south east}{\southwest \pgf@x=-\pgf@x}
  \anchor{text}{
    \pgfpointorigin
    \advance\pgf@x by -.5\wd\pgfnodeparttextbox%
    \advance\pgf@y by -.5\ht\pgfnodeparttextbox%
    \advance\pgf@y by +.5\dp\pgfnodeparttextbox%
  }

  \anchor{O}{ %
    \pgf@process{\northeast}%
    \pgf@x=0pt%
  }
  \anchor{P}{ %
    \pgf@process{\northeast}%
    \pgf@x=.5\pgf@x%
    \pgf@y=-\pgf@y%
  }
  \anchor{SA}{ %
    \pgf@process{\northeast}%
    \pgf@x=-.1\pgf@x%
    \pgf@y=-\pgf@y%
  }
  \anchor{SB}{ %
    \pgf@process{\northeast}%
    \pgf@x=-.3\pgf@x%
    \pgf@y=-\pgf@y%
  }
  \anchor{SC}{ %
    \pgf@process{\northeast}%
    \pgf@x=-.5\pgf@x%
    \pgf@y=-\pgf@y%
  }
  \anchor{SD}{ %
    \pgf@process{\northeast}%
    \pgf@x=-.7\pgf@x%
    \pgf@y=-\pgf@y%
  }
  \anchor{SE}{ %
    \pgf@process{\northeast}%
    \pgf@x=.9\pgf@x%
    \pgf@y=-\pgf@y%
  }
  \anchor{BOX}{ %
    \pgf@process{\northwest}%
    \pgf@x=\pgf@x%
    \pgf@y=-\pgf@y%
  }
  \anchor{LINE}{
    \pgf@process{\northeast}%
    \pgf@x=-1\pgf@x%
    \pgf@y=.\pgf@y%
  }
  \backgroundpath{
    \pgfpathrectanglecorners{\southwest}{\northeast}
    \pgf@anchor@spare@LINE
    \pgf@xa=\pgf@x \pgf@ya=\pgf@y
    \pgf@xb=\pgf@x \pgf@yb=\pgf@y
    \advance\pgf@xb by -2\pgf@x
    \pgfpathmoveto{\pgfpoint{\pgf@xa}{\pgf@ya}}
    \pgfpathlineto{\pgfpoint{\pgf@xb}{\pgf@ya}}
    \pgf@anchor@spare@BOX
    \pgf@xa=\pgf@x \pgf@ya=\pgf@y
    \pgf@xb=\pgf@x \pgf@yb=\pgf@y
    \pgf@xc=\pgf@x \pgf@yc=\pgf@y
    \advance\pgf@ya by -.\pgf@y
    \advance\pgf@xa by -1\pgf@x
    \advance\pgf@yb by -.8\pgf@y
    \advance\pgf@xb by -2\pgf@x
    \pgfpathmoveto{\pgfpoint{\pgf@xa}{\pgf@ya}}
    \pgfpathlineto{\pgfpoint{\pgf@xa}{\pgf@yb}}
    \pgfpathlineto{\pgfpoint{\pgf@xb}{\pgf@yb}}
    \pgfpathlineto{\pgfpoint{\pgf@xb}{\pgf@ya}}
    \pgfclosepath

    \begingroup
    \tikzset{spareM/port labels} %
    \tikz@textfont

    \pgf@anchor@spare@O
    \pgftext[top,at={\pgfpoint{\pgf@x}{\pgf@y}},y=-\pgfshapeinnerysep]{}

    \pgf@anchor@spare@P 
    \pgftext[bottom,at={\pgfpoint{\pgf@x}{\pgf@y}},y=\pgfshapeinnerysep]{}
    
    \pgf@anchor@spare@SA
    \pgftext[bottom,at={\pgfpoint{\pgf@x}{\pgf@y}},y=\pgfshapeinnerysep]{}
    
    \pgf@anchor@spare@SB
    \pgftext[bottom,at={\pgfpoint{\pgf@x}{\pgf@y}},y=\pgfshapeinnerysep]{}
    
    \pgf@anchor@spare@SC
    \pgftext[bottom,at={\pgfpoint{\pgf@x}{\pgf@y}},y=\pgfshapeinnerysep]{}
    
    \pgf@anchor@spare@SD
    \pgftext[bottom,at={\pgfpoint{\pgf@x}{\pgf@y}},y=\pgfshapeinnerysep]{}
    
    \pgf@anchor@spare@SE
    \pgftext[bottom,at={\pgfpoint{\pgf@x}{\pgf@y}},y=\pgfshapeinnerysep]{}
    \endgroup
  }
}

\tikzset{add font/.code={\expandafter\def\expandafter\tikz@textfont\expandafter{\tikz@textfont#1}}} 

\tikzset{spareM/port labels/.style={font=\sffamily\scriptsize}}
\tikzset{every spareM node/.style={draw,minimum height=1cm,minimum 
width=1.5cm,inner sep=1mm,outer sep=0pt,cap=round,add 
font=\sffamily}}

\pgfdeclareshape{fdep}{
  \savedanchor\northeast{%
    \pgfmathsetlength\pgf@x{\pgfshapeminwidth}%
    \pgfmathsetlength\pgf@y{\pgfshapeminheight}%
    \pgf@x=0.5\pgf@x
    \pgf@y=0.5\pgf@y
  }
  \savedanchor\southwest{%
    \pgfmathsetlength\pgf@x{\pgfshapeminwidth}%
    \pgfmathsetlength\pgf@y{\pgfshapeminheight}%
    \pgf@x=-0.5\pgf@x
    \pgf@y=-0.5\pgf@y
  }
  \inheritanchorborder[from=rectangle]

  \anchor{center}{\pgfpointorigin}
  \anchor{north}{\northeast \pgf@x=0pt}
  \anchor{east}{\northeast \pgf@y=0pt}
  \anchor{south}{\southwest \pgf@x=0pt}
  \anchor{west}{\southwest \pgf@y=0pt}
  \anchor{north east}{\northeast}
  \anchor{north west}{\northeast \pgf@x=-\pgf@x}
  \anchor{south west}{\southwest}
  \anchor{south east}{\southwest \pgf@x=-\pgf@x}
  \anchor{text}{
    \pgfpointorigin
    \advance\pgf@x by -.5\wd\pgfnodeparttextbox%
    \advance\pgf@y by -.5\ht\pgfnodeparttextbox%
    \advance\pgf@y by +.5\dp\pgfnodeparttextbox%
  }

  \anchor{O}{ %
    \pgf@process{\northeast}%
    \pgf@x=0pt%
  }
  \anchor{EA}{ %
    \pgf@process{\northeast}%
    \pgf@x=.1\pgf@x%
    \pgf@y=-\pgf@y%
  }
  \anchor{EB}{ %
    \pgf@process{\northeast}%
    \pgf@x=.3\pgf@x%
    \pgf@y=-\pgf@y%
  }
  \anchor{EC}{ %
    \pgf@process{\northeast}%
    \pgf@x=.5\pgf@x%
    \pgf@y=-\pgf@y%
  }
  \anchor{ED}{ %
    \pgf@process{\northeast}%
    \pgf@x=.7\pgf@x%
    \pgf@y=-\pgf@y%
  }
  \anchor{EE}{ %
    \pgf@process{\northeast}%
    \pgf@x=.9\pgf@x%
    \pgf@y=-\pgf@y%
  }
  \anchor{T}{ %
    \pgf@process{\northeast}%
    \pgf@x=-1\pgf@x%
    \pgf@y=-.5\pgf@y%
  }
  \anchor{LINE}{
    \pgf@process{\northeast}%
    \pgf@x=-1\pgf@x%
    \pgf@y=.\pgf@y%
  }
  \backgroundpath{
    \pgfpathrectanglecorners{\southwest}{\northeast}
    \pgf@anchor@fdep@LINE
    \pgf@xa=\pgf@x \pgf@ya=\pgf@y
    \pgf@xb=\pgf@x \pgf@yb=\pgf@y
    \advance\pgf@xb by -2\pgf@x
    \pgfpathmoveto{\pgfpoint{\pgf@xa}{\pgf@ya}}
    \pgfpathlineto{\pgfpoint{\pgf@xb}{\pgf@ya}}
    \pgf@anchor@fdep@T
    \pgf@xa=\pgf@x \pgf@ya=\pgf@y
    \pgf@xb=\pgf@x \pgf@yb=\pgf@y
    \pgf@xc=\pgf@x \pgf@yc=\pgf@y
    \advance\pgf@ya by -.3333\pgf@x
    \advance\pgf@xb by -.5\pgf@x
    \advance\pgf@yb by .\pgf@x
    \advance\pgf@yc by .3333\pgf@x
    \pgfpathmoveto{\pgfpoint{\pgf@xa}{\pgf@ya}}
    \pgfpathlineto{\pgfpoint{\pgf@xb}{\pgf@yb}}
    \pgfpathlineto{\pgfpoint{\pgf@xc}{\pgf@yc}}
    \pgfclosepath

    \begingroup
    \tikzset{fdep/port labels} %
    \tikz@textfont

    \pgf@anchor@fdep@O
    \pgftext[top,at={\pgfpoint{\pgf@x}{\pgf@y}},y=-\pgfshapeinnerysep]{}
    
    \pgf@anchor@fdep@EA
    \pgftext[bottom,at={\pgfpoint{\pgf@x}{\pgf@y}},y=\pgfshapeinnerysep]{}
    
    \pgf@anchor@fdep@EB
    \pgftext[bottom,at={\pgfpoint{\pgf@x}{\pgf@y}},y=\pgfshapeinnerysep]{}
    
    \pgf@anchor@fdep@EC
    \pgftext[bottom,at={\pgfpoint{\pgf@x}{\pgf@y}},y=\pgfshapeinnerysep]{}
    
    \pgf@anchor@fdep@ED
    \pgftext[bottom,at={\pgfpoint{\pgf@x}{\pgf@y}},y=\pgfshapeinnerysep]{}
    
    \pgf@anchor@fdep@EE
    \pgftext[bottom,at={\pgfpoint{\pgf@x}{\pgf@y}},y=\pgfshapeinnerysep]{}
    \endgroup
  }
}

\tikzset{add font/.code={\expandafter\def\expandafter\tikz@textfont\expandafter{\tikz@textfont#1}}} 

\tikzset{fdep/port labels/.style={font=\sffamily\scriptsize}}
\tikzset{every fdep node/.style={draw,minimum height=1cm,minimum 
width=1.5cm,inner sep=1mm,outer sep=0pt,cap=round,add 
font=\sffamily}}

\makeatother

\tikzstyle{every picture}=[>=stealth]
\tikzstyle{and}=[and gate US,draw,rotate=90,logic gate inputs=nnn,minimum width=1cm,anchor=east,xshift=-1mm]
\tikzstyle{or}=[or gate US,draw,rotate=90,logic gate inputs=nnn,minimum width=1cm,anchor=east,xshift=-1mm]
\tikzstyle{be}=[circle,draw,fill=white!100,anchor=north, minimum width=0.5cm]
\tikzstyle{triangle}=[isosceles triangle,draw,shape border rotate=90,isosceles triangle stretches=true, minimum height=0.6cm,minimum width=0.6cm]

\tikzset{or1/.style={or gate US,fill=black!10,draw,rotate=90, scale=1.25}}
\tikzset{and1/.style={fill=black!10,and gate US,draw,rotate=90, scale=1.25}}
\tikzset{be/.style={circle,fill=black!20,draw, minimum size=3cm}}
\tikzset{triangle/.style={isosceles triangle,draw,shape border rotate=90,isosceles triangle stretches=true,minimum width=0.8cm,minimum height=0.8cm}}